 \documentclass[10pt, final]{IEEEtran}

\usepackage{amsthm}

\usepackage{amssymb}
\usepackage{amsmath}
\usepackage{bbm}
\usepackage{mathrsfs}
\usepackage{cite}
\usepackage{bm,epsfig,amsthm,url}
\usepackage{indentfirst}
\usepackage{amsfonts}
\usepackage{epstopdf}
\usepackage{cases}
\usepackage[caption=false,font=scriptsize]{subfig}
\usepackage{xcolor}
\usepackage{mathtools}

\makeatletter
\renewcommand{\maketag@@@}[1]{\hbox{\m@th\normalsize\normalfont#1}}%
\makeatother
\usepackage{hyperref}
\usepackage{stfloats}

\newtheoremstyle{mystyle}{}{}{}{}{}{: }{0pt}{\indent \it{\thmname{#1}\thmnumber{ #2}\thmnote{#3}}}
\theoremstyle{mystyle}

\newtheorem{Proposition}{Proposition}

\usepackage[noend]{algpseudocode}
\usepackage{float}
\usepackage{algorithmicx}
\usepackage[linesnumbered,ruled,vlined]{algorithm2e}

\begin{document}

\title{Rotatable IRS Aided Wireless Communication}

\author{{Qiaoyan~Peng,~\IEEEmembership{Graduated~Student~Member,~IEEE},~Qingqing~Wu,~\IEEEmembership{Senior Member,~IEEE},~Guangji~Chen,~Wen~Chen,~\IEEEmembership{Senior Member,~IEEE},~Shaodan~Ma,~\IEEEmembership{Senior Member,~IEEE},~Shanpu~Shen,~\IEEEmembership{Senior Member,~IEEE},~Rui~Zhang,~\IEEEmembership{Fellow,~IEEE}}
\thanks{Q. Peng is with the Department of Electronic Engineering, Shanghai Jiao Tong University, Shanghai 200240, China, and also with the State Key Laboratory of Internet of Things for Smart City, University of Macau, Macao 999078, China (email: qiaoyan.peng@connect.um.edu.mo). 
	Q. Wu and W. Chen are with the Department of Electronic Engineering, Shanghai Jiao Tong University, Shanghai 200240, China (email: qingqingwu@sjtu.edu.cn; wenchen@sjtu.edu.cn).
	G. Chen is with Nanjing University of Science and Technology, Nanjing 210094, China (email: guangjichen@njust.edu.cn).
	S. Ma and S. Shen are with the State Key Laboratory of Internet of Things for Smart City, University of Macau, Macao 999078, China (email: shaodanma@um.edu.mo; shanpushen@um.edu.mo).
	R. Zhang is with the Department of Electrical and Computer Engineering, National University of Singapore, Singapore 117583 (e-mail: elezhang@nus.edu.sg).
}
}

\maketitle

\begin{abstract}
Rotatable intelligent reflecting surface (IRS) introduces a new spatial degree of freedom (DoF) by dynamically adjusting orientations without the need of changing its elements' positions in real time. To unleash the full potential of rotatable IRSs for wireless communications, this paper investigates the joint optimization of IRS rotation angles to maximize the minimum expected signal-to-noise ratio (SNR) over all locations within a given target area. We first propose an angle-dependent channel model that accurately characterizes the reception and reflection of each IRS element. Different from the conventional cosine-law assumption, the proposed model captures the practical electromagnetic characteristics of the IRS, including the effective reception area and reflection efficiency. For the single target location case, a particle swarm optimization (PSO)-based algorithm is developed to solve the SNR maximization problem, and a closed‑form expression for a near‑optimal solution is derived to provide useful insights. For the general area coverage enhancement case, the optimal rotation is obtained through a two‑loop PSO‑based iterative algorithm with null‑point detection.
In this algorithm, the outer loop updates the global rotation angles to maximize the minimum SNR over the target area, whereas the inner loop evaluates the SNR distribution within the area to identify the location corresponding to the minimum SNR through null‑point detection. Numerical results demonstrate significant SNR improvement achieved by the proposed rotatable IRS design over various benchmark schemes under different system setups.
\end{abstract}
\begin{IEEEkeywords}
Intelligent reflecting surface (IRS), rotatable IRS, rotation optimization, particle swarm optimization (PSO), coverage enhancement.
\end{IEEEkeywords}

\section{Introduction}
The development of the sixth-generation (6G) wireless networks is driven by the ambition to achieve unprecedented data rates, superior energy efficiency, and broader network coverage, which aim to support the demanding requirements of future ubiquitous connectivity \cite{wu_6G,AIRS_magazine,xue_survey}. As a promising solution, intelligent reflecting surfaces (IRSs) have been introduced as a cost- and energy-efficient complement to traditional wireless networks, offering proactive and flexible control of the radio-wave propagation \cite{9606864,9650755,9652042}. By dynamically adjusting the amplitude and/or phase of passive elements, IRSs can reconfigure wireless channels to strengthen desired signals, suppress interference, improve physical-layer security, and extend network coverage in challenging propagation scenarios. These capabilities allow the IRS to support diverse applications such as interference mitigation in dense networks, throughput improvement for cell-edge users, and simultaneous wireless information and power transfer for energy-limited devices \cite{CI,ISI,chen_iot}. Motivated by these prospects, extensive research has been conducted on IRS-aided systems, focusing on main issues including energy efficiency maximization, joint active and passive beamforming design, and the theoretical analysis of achievable rate regions \cite{hirs_peng,kang_deployment,fixed_feng,fu_deployment}.

The aforementioned studies mainly consider IRSs that are installed at fixed locations with pre-determined orientations to enhance wireless communication systems. In existing works, the IRS is primarily deployed in terrestrial scenarios, where it is typically mounted on the outer walls of buildings, indoor ceilings, and other static structures \cite{double_IRS,peng_deployment,magazine}. As a fixed IRS can only provide effective service to transmitters and receivers located within the same half-space relative to its reflective surface, communication nodes positioned on the opposite sides of the IRS cannot benefit from its reflection capability, which restricts its spatial coverage. Moreover, although passive beamforming can be used to reflect signals toward users, each reflection may lead to partial scattering of the electromagnetic waves in undesired directions. This dispersion of signal energy results in additional propagation loss and severe attenuation, which significantly degrade the overall signal quality and reduce the efficiency of the IRS-aided links, especially in complex propagation environments.

To address the limitations of conventional fixed IRSs, recent research has explored dynamic deployment strategies that can operate in a more adaptive and flexible manner. These strategies introduce movable and rotatable IRS in wireless communication systems, thereby opening new opportunities for fully harnessing the potential of the IRS technology in next-generation wireless systems. To be specific, aerial IRS–enabled three-dimensional (3D) wireless networks, where the IRS is mounted on aerial platforms, e.g., unmanned aerial vehicles (UAVs), have received significant attention for intelligent signal reflection \cite{lu_distance,liu_uav,uav_jiang,uav_wang}. Different from traditional terrestrial IRSs, which are inherently restricted to serving only the nodes located within the same half-space of the reflective surface, an aerial IRS can achieve near-omnidirectional reflection and facilitate line-of-sight (LoS) communication between transmitters and receivers due to its elevated position relative to the ground. In \cite{lu_distance}, transmit beamforming, IRS placement, and 3D passive beamforming were jointly optimized to maximize the worst-case signal-to-noise ratio (SNR) within a target area, utilizing a beam broadening and flattening technique to achieve significant performance gains over the fixed IRS. Furthermore, the maneuverability of UAVs allows the location and orientation of an aerial IRS to be dynamically adjusted, which introduces additional degrees of freedom for system optimization and enables highly flexible 3D network design that can adapt to real-time changes in communication demands. The work \cite{liu_uav} developed an efficient alternating optimization algorithm in a UAV-mounted IRS-aided multicast system with the joint design of the IRS's 3D location, 3D orientation, and passive beamforming. Despite many advantages mentioned above, aerial IRS poses fundamental limitations for several reasons. First, the endurance, stability, and controllability of the aerial platform remain critical technical concerns, as the platform must support sustained operation under varying environmental conditions \cite{endurance}. Second, platform drift and vibration, which influence reflection accuracy and system stability, as well as the complexity associated with optimizing the placement and trajectory of aerial platforms, impose significant design burdens \cite{complexity}. Third, regulatory constraints related to flight zones and airspace management may further limit the large-scale or long-term deployment of aerial IRSs in practical scenarios \cite{no-fly}. 

Considering the ease of implementation provided by the fixed IRS and the high flexibility offered by the aerial IRS, the rotatable IRS has been proposed, which focuses on dynamic orientation adjustment rather than full spatial displacement \cite{rotate4,rotate1,rotate2,rotate3,rotate7,6DMA1}. To be specific, a rotatable IRS can flexibly vary its rotation angles at a fixed position, thereby dynamically reconfiguring its reception and reflection patterns and providing additional spatial degrees of freedom without physical movement. By properly adjusting its rotation angle, a single rotatable IRS can achieve panoramic coverage or full-angle reflection, enabling it to serve any transmitter-receiver pair within its coverage region. Moreover, the practical implementation of a rotatable IRS resembles existing rotatable antennas, where the IRS is mounted on a motor-driven shaft, and its rotation angles is precisely controlled through mechanical adjustment mechanisms, which can also be considered as a special case of the general six-dimensional movable antenna (6DMA) model with both 3D position and 3D orientation control \cite{6DMA2,6DMA3, 6DMA4, RA_zheng,FR_lu,ning}. Such an architecture ensures strong compatibility with existing wireless infrastructures while avoiding the need for continuous positioning, complex mechanical control, and the high power consumption typically associated with mobile platforms. Because of its simple structure, low energy cost, and low latency, a rotatable IRS is appealing for scenarios that require rapid adaptation but where large-scale mechanical movement is impractical. The rotatable IRS has been extensively investigated under various communication systems, including cellular systems \cite{rotate1}, mobile edge computing systems \cite{rotate2}, and multicast systems \cite{rotate3}. In \cite{rotate1}, the received power was maximized when the incident angle was zero. The work \cite{rotate2} proposed a deep learning-based algorithm for the joint design of IRS orientation, phase shifts, and offloading strategies. The base station (BS) beamforming, IRS phase shifts, and orientation adjustment were jointly optimized to maximize the multicast rates in \cite{rotate2}. 

Despite the effectiveness of the IRS rotation strategies, most existing studies have relied on a simplified and idealized channel model that primarily accounts for the propagation path loss and orientation parameters, while overlooking other crucial physical-layer characteristics such as the effective reception area of the IRS, as well as its reception and reflection efficiency \cite{physical1,physical2,physical3}. Consequently, the corresponding deployment strategies may underperform in practical scenarios with non-ideal channels, as they fail to capture the geometric and physical constraints inherent in practical channel conditions \cite{PL}. To address this issue, the work \cite{rotate5} assumed an IRS with predetermined electromagnetic properties and focused on how its steering capability affects the spatial power redistribution of the reflected beam. Furthermore, a physics‑based model was proposed in \cite{rotate6}, where the IRS elements were grouped into tiles whose effects on the wireless channel were characterized by solving the electric and magnetic field integral equations. In the above studies, the IRS is represented as a continuous reflecting surface whose global response determines the scattered beam. It provides analytical simplicity and facilitates theoretical investigations. In contrast, the IRS can be modeled as a planar distribution of discrete controllable elements that capture the collective response produced by the individual reflecting elements, which provides a more accurate and practical framework for IRS design by capturing its physical characteristics. One of the main advantages lies in allowing the electromagnetic properties to be directly integrated with circuit‑level models of the tunable elements, thereby bridging the gap between idealized analysis and practical design. In \cite{cos1}, the authors evaluated the scattering performance, power, and size of a single IRS element and then maximized the received signal power toward a specific user. Furthermore, the rotation of the IRS was optimized to maximize the worst SNR over all locations in a target area in both indoor \cite{cos2} and outdoor scenarios \cite{cos3}. However, in \cite{cos1,cos2,cos3}, the radiation patterns of the transmit and receive antennas, as well as the IRS elements, are modeled with angle‑dependent factors derived from a simplified cosine law, which may lead to limitations of rotation optimization in terms of SNR. 


Motivated by the above discussion, this paper investigates a rotatable IRS-aided wireless communication system, where the IRS is deployed to extend BS coverage toward a specific target area, e.g., a cellular hot spot or a remote zone lacking access. Specifically, we aim to design the IRS rotation to maximize the worst-case/minimum SNR in the target area, thereby improving the performance. The corresponding optimization problem is challenging to solve directly for the following reasons. First, in contrast to the cosine law-based model, a more practical channel model is required to capture the impact of the IRS rotation by characterizing the reception and reflection behaviors of each IRS element, which lays the foundation for further performance analysis. Since this elaborate model introduces additional spatial dependencies and nonlinear coupling into the problem formulation, it significantly increases the analytical complexity. Second, different from the conventional fixed IRS, the rotation of the rotatable IRS introduces a new set of design variables that influence not only the incident and reflection angles at the IRS but also the angular extent from the IRS that covers the target area. These coupled effects determine the effective coverage range and directivity of the reflected beam, thereby significantly affecting the quality of the received signals. Third, compared with prior studies that steer the IRS beam toward a single user at a known location, the IRS rotation is optimized by taking into account all possible user positions within a given target area, which leads to a more complex and challenging optimization problem. The main contributions of this paper are summarized as follows:

\begin{itemize}
	\item First, we study a rotatable IRS-aided wireless communication system by optimizing the rotation angles of the IRS. A practical angle-dependent channel model based on electromagnetic field modeling is proposed, which accounts for the reception and reflection factors. With the optimized BS and IRS beamforming for each user location, we derive the closed-form expression of the corresponding received SNR with respect to (w.r.t.) the IRS rotation angle. By utilizing the proposed model, we formulate an optimization problem aimed at maximizing the minimum SNR over all locations in the given area, subject to the rotation constraint to ensure effective reflection.
	\item Next, we consider the case of SNR maximization at a specific user location and propose an efficient particle swarm optimization (PSO)-based algorithm for the IRS rotation design. Given that the rotatable IRS is generally electrically small, we derive a closed-form expression for a near-optimal solution. Furthermore, the optimal elevation angle approaches $-\pi/2$ when the IRS is deployed at a sufficiently high altitude, which is independent of the azimuth angle. Moreover, no rotation is needed when both the BS and the user are far away from the IRS along the $x$-axis (see Fig. \ref{fig:systemmodel}). For the general case of min-SNR maximization in a target area, we propose a PSO-based two-loop Iterative algorithm. To reduce the complexity, null point detection can be applied to avoid exhaustive search over all user locations.
	\item Finally, extensive numerical results are presented to validate our theoretical results and demonstrate the significant performance gains achieved by the rotatable IRS over other benchmark schemes under different setups. Moreover, it is shown that our proposed scheme achieves about 3 dB gain in SNR over conventional fixed IRS scheme for the special case of a single target location.
\end{itemize}

The remainder of this paper is organized as follows. Section \ref{System Model} introduces the system model and formulates the optimization problem for the rotatable IRS-aided wireless communication system. Sections \ref{Single Target Location SNR Optimization} and \ref{Area Coverage Enhancement} propose efficient algorithms for the single target location and area coverage scenarios, respectively. Section \ref{Simulation Results} provides numerical results that evaluate our proposed designs. Section \ref{Conclusion} concludes the paper.

Notations: Scalars are represented by italic letters. Vectors and matrices are denoted by bold lowercase and uppercase letters, respectively. The modulo operation $\operatorname{mod}\left(a,b\right)$ returns the remainder of the division of $a$ by $b$. For real number $s$, $\left\lfloor s \right\rfloor$ denotes the floor operation and $\operatorname{sinc}\left(s\right) \triangleq \sin \left(s\right)/s$. For a vector $\mathbf{s}$, $\left\|\mathbf{s}\right\|$, and $\left[\mathbf{s}\right]_n$ denote its Euclidean norm and $n$-th entry, respectively. For a matrix $\mathbf{S}$, $\left[\mathbf{S}\right]_{m,n}$ denotes the element on the $m$-th row and $n$-th column of $\mathbf{S}$. $j$ denotes the imaginary unit with $j^2 = -1$.

\section{System Model}
\label{System Model}
\begin{figure}[t]
	\centering
	\includegraphics[width=0.35\textwidth]{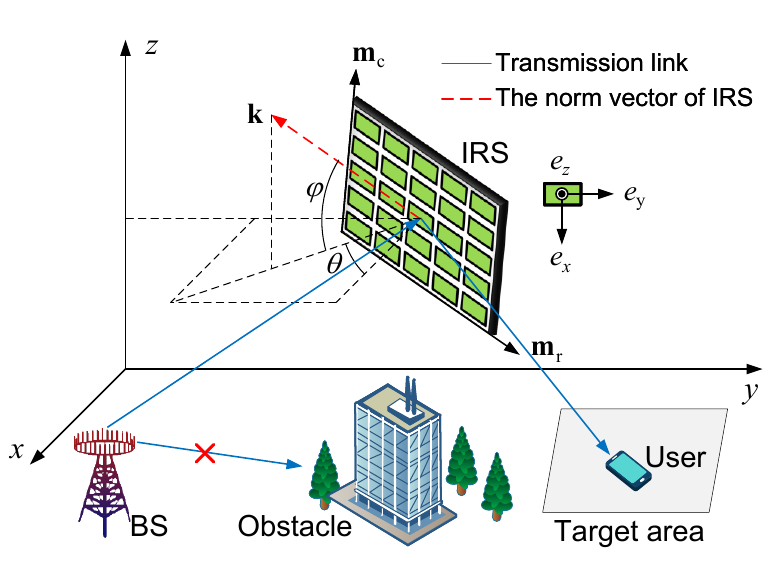}
	\vspace{-10pt}
	\caption{A rotatable IRS-aided wireless communication system.}
	\label{fig:systemmodel}
\end{figure}
As shown in Fig. \ref{fig:systemmodel}, we consider an IRS-aided wireless communication system, where the IRS is deployed to assist from the multi-antenna BS to the single-antenna user within the target area $\mathcal{A}$. We focus on a challenging scenario where the direct BS-user link is blocked by obstacles and assume that all the considered channels follow the free-space LoS model \cite{LOS1,LOS2}. The IRS consists of a total of $N = N_{\mathrm{c}} \times N_{\mathrm{r}}$ elements, arranged in $N_{\mathrm{r}}$ rows and $N_{\mathrm{c}}$ columns, with each element have the size of $\bar{l} \times \bar{l}$. The BS is equipped with a conventional UPA placed on the $x$-$y$ plane, and the total number of antennas is $M = M_\mathrm{c} \times M_\mathrm{r}$ antennas, where $M_\mathrm{r}$ and $M_\mathrm{c}$ denote the number of antennas along the $x$- and $y$-axis, respectively. In the considered system, we establish a global Cartesian coordinate system (CCS). Assuming that the target area $\mathcal{A}$ is a rectangular area with length $A_\mathrm{x}$ and width $A_\mathrm{y}$ on the $x$-$y$ plane, its center is denoted as $\bm{p}_\mathrm{r} = \left[x_\mathrm{r},y_\mathrm{r},0\right]^T$. Thus, any user location within the target area $\mathcal{A}$ is located at $\bm{p}_\mathrm{U} = \left[x_\mathrm{U},y_\mathrm{U},0\right]^T$ with $x_\mathrm{U} \in \left[x_\mathrm{r} - A_\mathrm{x}/2, x_\mathrm{r} + A_\mathrm{x}/2\right]$ and $y_\mathrm{U} \in \left[y_\mathrm{r} - A_\mathrm{y}/2, y_\mathrm{r} + A_\mathrm{y}/2\right]$. 

Let $\bm{\Omega} = (\theta, \phi)^T$ denote the rotation angles of the IRS, where $\theta$ and $\varphi$ are the azimuth and elevation angles, respectively, within the feasible set $\mathcal{S} = \left\{(\theta, \varphi)^T|\theta \in \left[-\pi/2,\pi/2\right], \varphi \in \left[-\pi/2,\pi/2\right]\right\}$. Thus, the normal vector of the IRS is expressed as
\begin{align}
	{{\bm{k}}} = {\left( {\cos {\theta}\cos {\varphi}, - \sin {\theta}\cos {\varphi},\sin {\varphi}} \right)^T}.
\end{align}
The reflecting elements of the IRS are placed on lines along two orthogonal base directions, which are given by
\begin{align}
	{\bm{m}}_{\mathrm{c}} &= {\left( { - \cos {\theta} \sin {\varphi},\sin {\theta}\sin {\varphi},\cos {\varphi}} \right)^T},\\
	{\bm{m}}_{\mathrm{r}} &= {\left( {\sin {\theta},\cos {\theta},0} \right)^T}.
\end{align}
In the initial stage, the IRS is set without rotation, i.e., $\theta = 0$ and $\varphi = 0$. As such, we have ${{\bm{k}}} = {\left( 1,0,0 \right)^T}$, $
{\bm{m}}_{\mathrm{r}} = {\left( {0,1,0} \right)^T}$ and ${\bm{m}}_{\mathrm{c}} = {\left( 0,0,1 \right)^T}$, which are the unit vectors along the $x$-, $y$-, and $z$-axis of the global CCS, respectively.

\begin{figure}[t]
	\centering
	\includegraphics[width=0.25\textwidth]{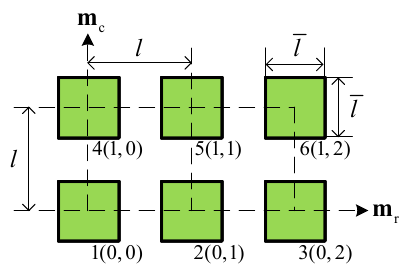}
	\vspace{-5pt}
	\caption{An example of a rectangular IRS with 6 elements, where each reflecting element in a specific row and a specific column is labeled with a unique index.}
	\label{fig:mapmodel}
	\vspace{-10pt}
\end{figure}

As illustrated in Fig. \ref{fig:mapmodel}, without loss of generality, the IRS reflecting elements are assumed to be arranged in a rectangular array. The elements are arranged along lines in the two base directions, i.e., ${\bm{m}}_{\mathrm{r}}$ and ${\bm{m}}_{\mathrm{c}}$, with an inter-element spacing of $l$ in each direction. We map the reflecting element located at $N_{n}^\mathrm{r}$-th row and $N_{n}^\mathrm{c}$-th column as the $n$-th element of the IRS, with $N_{n}^\mathrm{r} \in \mathcal{N_\mathrm{r}} = \{0,\cdots,N_\mathrm{r}-1\}$, $N_{n}^\mathrm{c} \in \mathcal{N_\mathrm{c}} = \{0,\cdots,N_\mathrm{c}-1\}$, and $n \in \mathcal{N} = \{1,\cdots,N\}$, where 
\begin{align}
	N_{{n}}^{\mathrm{r}} &= \left\lfloor {\left( {{n} - 1} \right)/N_{\mathrm{c}}} \right\rfloor, n \in \mathcal{N},\\
	N_{{n}}^{\mathrm{c}} &= {n} - N_{\mathrm{c}} N_{{n}}^{\mathrm{r}} - 1,n \in \mathcal{N}.
\end{align}
The bottom-left element of the IRS is defined as the 1st element of the UPA whose coordinate is given by ${{\bm{p}}_{1}} = {\left[ {{x_1},{y_1},{z_1}} \right]^T}$. We take the 1st element of the IRS as the reference point. Accordingly, the coordinate for the $n$-th element of the IRS is expressed as
\begin{align}
	\label{p_n}
	{{\bm{p}}_{{n}}} = {{\bm{p}}_{1}} + N_{{n}}^{\mathrm{c}}l{\bm{m}}_{\mathrm{r}} + N_{{n}}^{\mathrm{r}}l{\bm{m}}_{\mathrm{c}},n \in \mathcal{N},	
\end{align}
where $l$ is the distance between two adjacent IRS elements. Then, the center coordinate for the IRS is expressed as 
\begin{align}
	{{\bm{p}}_{{\mathrm{c}}}} = {{\bm{p}}_{1}} + \frac{{N_{\mathrm{c}} - 1}}{2}l{\bm{m}}_{\mathrm{r}} + \frac{{N_{\mathrm{r}} - 1}}{2}l{\bm{m}}_{\mathrm{c}},n \in \mathcal{N}.
\end{align}
Similar to the IRS, we map the antenna located at $M_{m}^\mathrm{r}$-th row and $M_{m}^\mathrm{c}$-th column as the $m$-th antenna of the BS, with $M_{m}^\mathrm{r} \in \mathcal{M_\mathrm{r}} = \{0,\cdots,M_\mathrm{r}-1\}$, $M_{m}^\mathrm{c} \in \mathcal{M_\mathrm{c}} = \{0,\cdots,M_\mathrm{c}-1\}$, and $m \in \mathcal{M} = \{1,\cdots,M\}$, where 
\begin{align}
	M_{m}^{\mathrm{r}} &= \left\lfloor {\left( {{m} - 1} \right)/M_{\mathrm{c}}} \right\rfloor, m \in \mathcal{M},\\
	M_{m}^{\mathrm{c}} &= {m} - M_{\mathrm{c}} M_{{m}}^{\mathrm{r}} - 1, m \in \mathcal{M}.
\end{align}
We denote the coordinates for the 1st antenna of the BS as ${{\bm{p}}_{\mathrm{B},1}} = {\left[{{x_{\mathrm{B,1}}},{y_{\mathrm{B},1}},0} \right]^T}$. Accordingly, the coordinate for the $m$-th antenna of the BS is given by 
\begin{align}
	\label{p_b,m}
	{{\bm{p}}_{\mathrm{B},m}} = {{\bm{p}}_{{\mathrm{B}},1}} + M_{m}^{\mathrm{c}}\tilde l{{\bm{m}}_{\mathrm{B}}^{\mathrm{c}}} + M_{m}^{\mathrm{r}}\tilde l{{\bm{m}}_{\mathrm{B}}^{\mathrm{r}}}, m \in \mathcal{M}.
\end{align}
where $\bm{m}_\mathrm{B}^{\mathrm{r}} = \left[0,1,0\right]^T$, $\bm{m}_\mathrm{B}^{\mathrm{c}} = \left[-1,0,0\right]^T$, and $\tilde l$ is the distance between two adjacent BS antennas. Then, the center coordinate for the BS is expressed as 
\begin{align}
	{{\bm{p}}_{{\mathrm{B}}}} = {{\bm{p}}_{{\mathrm{B}},1}} + \frac{{M_{\mathrm{c}} - 1}}{2}\tilde l {{\bm{m}}_{\mathrm{B}}^{\mathrm{c}}} + \frac{{M_{\mathrm{r}} - 1}}{2}\tilde l {{\bm{m}}_{\mathrm{B}}^{\mathrm{r}}}.
\end{align}
The distance between the center of the BS and the center of the IRS, as well as the distance between the center of the IRS and any location of the target area ${{\bm{p}}_{\mathrm{U}}} \in \mathcal{A}$ are given by
\begin{align}
	t = \left\| {{{\bm{p}}_{\mathrm{B}}} - {{\bm{p}}_{\mathrm{c}}}} \right\|, \;\; r = \left\| {{{\bm{p}}_{\mathrm{U}}} - {{\bm{p}}_{\mathrm{c}}}} \right\|,
\end{align}
respectively. Under the LoS channel, the entry in column $n$ of the channel from the IRS to the user, i.e., $\bm{f}^H$, is given by
\begin{align}
	\left[\bm{f}^H\right]_{n} = \frac{{\sqrt \beta }}{{{r_{{n}}}}}{e^{ - j\frac{{2\pi }}{\lambda }{r_{{n}}}}},n \in \mathcal{N},
\end{align}
where ${r_{{n}}} = \left\| {{{\bm{p}}_{\mathrm{U}}} - {{\bm{p}}_{n}}} \right\|$ is the distance between the user and the $n$-th element of the IRS, $\beta$ is the channel power gain at the reference distance, i.e., 1 meter (m), and $\lambda$ denotes the carrier wavelength.

The entry in row $n$ and column $m$ of the channel from the $m$-th antenna of the BS to the $n$-th element of the IRS, i.e., $\bm{G}$, is given by
\begin{align}
	\label{G}
	\left[\bm{G}\right]_{n,m} = \frac{{\sqrt \beta }}{{{t_{{n,m}}}}}{e^{ - j\frac{{2\pi }}{\lambda }{t_{{n,m}}}}},n \in \mathcal{N},m \in \mathcal{M},
\end{align}
where $t_{n,m} = \left\| {{{\bm{p}}_{\mathrm{B},m}} - {{\bm{p}}_{n}}} \right\|$ is the distance from the $m$-th antenna of the BS to the $n$-th element of the IRS.

The received signal at the typical user within the target area is given by
\begin{align}
	y = {{\bm{f}}^H}{{\bm{\Phi }}}{\bm{Gw}}s + n_0, 
\end{align}
where the coefficient matrix of the IRS is $\bm{\Phi} = \operatorname{diag} \left(\eta_{1}, \cdots, \eta_{N}\right)$ with the reflection coefficients $\eta_n,n \in \mathcal{N}$, $\bm{w} = \left[w_1, \cdots, w_M\right]^T$ denotes the BS beamforming vector satisfying $\|\bm{w}\|^2 \le P_\mathrm{t}$, $s$ stands for the symbol transmitted from BS with power $P_\mathrm{t}$, $n_0 \sim \mathcal{CN} \left(0,\sigma_0^2\right)$ denotes the additive Gaussian white noise (AWGN) at the receiver with $\sigma_0^2$ representing the noise power. 

We denote $\varpi _{\mathrm{B}}^{\mathrm{c}} = \arccos \left( {\frac{{( {{{\bm{p}}_{{\mathrm{B,1}}}} - {{\bm{p}}_1}}) \cdot {\bm{m}}_{\mathrm{B}}^{\mathrm{c}}}}{{\| {{{\bm{p}}_{{\mathrm{B,1}}}} - {{\bm{p}}_1}} \| \cdot \| {{\bm{m}}_{\mathrm{B}}^{\mathrm{c}}} \|}}} \right)$, $\varpi _{\mathrm{B}}^{\mathrm{r}} = \arccos \left( {\frac{{( {{{\bm{p}}_{{\mathrm{B,1}}}} - {{\bm{p}}_1}} ) \cdot {\bm{m}}_{\mathrm{B}}^{\mathrm{r}}}}{{\| {{{\bm{p}}_{{\mathrm{B,1}}}} - {{\bm{p}}_1}} \| \cdot \| {{\bm{m}}_{\mathrm{B}}^{\mathrm{r}}} \|}}} \right)$, ${\varpi _{\mathrm{c}}} = \arccos \left( {\frac{{( {{{\bm{p}}_{{\mathrm{B,1}}}} - {{\bm{p}}_1}} ) \cdot {\bm{m}}_{\mathrm{c}}}}{{\| {{{\bm{p}}_{{\mathrm{B,1}}}} - {{\bm{p}}_1}} \| \cdot \| {{\bm{m}}_{\mathrm{c}}} \|}}} \right)$, and ${\varpi _{\mathrm{r}}} = \arccos \left( {\frac{{( {{{\bm{p}}_{{\mathrm{B,1}}}} - {{\bm{p}}_1}} ) \cdot {\bm{m}}_{\mathrm{r}}}}{{\| {{{\bm{p}}_{{\mathrm{B,1}}}} - {{\bm{p}}_1}} \| \cdot \| {{\bm{m}}_{\mathrm{r}}} \|}}} \right)$ as the angles between $({{{\bm{p}}_{{\mathrm{B,1}}}} - {{\bm{p}}_1}})$ and ${{\bm{m}}_{\mathrm{B}}^{\mathrm{c}}}$, ${{\bm{m}}_{\mathrm{B}}^{\mathrm{r}}}$, ${{{\bm{m}}_{\mathrm{c}}}}$, and ${{{\bm{m}}_{\mathrm{r}}}}$, respectively. According to \cite{double_IRS}, the LoS channel from the BS to the IRS can be assumed to be a rank-one matrix if $t_{1,1} \ge \sqrt{N}l^2/\lambda$. To facilitate the design of the BS and IRS beamforming, we provide a tractable characterization for the BS-IRS channel in the following proposition. 
\begin{Proposition}
	\label{pro:1}
	Under the assumption that $t_{1,1} \ge \sqrt{N}l^2/\lambda$, the BS-IRS channel entry can be approximated as
	\begin{align}
		{\left[ {\bm{G}} \right]_{n,m}} \approx \frac{{\sqrt \beta }}{{{t_{n,m}}}}{e^{ - j\frac{{2\pi }}{\lambda }\left( {{t_{1,1}} + {{{{\bar g}}}_m} + {{{{\tilde g}}}_n}} \right)}},n \in \mathcal{N},m \in \mathcal{M},
	\end{align}
	where ${{{{\bar g}}}_m} = M_m^{\mathrm{c}}\tilde l\cos \varpi _{\mathrm{B}}^{\mathrm{c}} + M_m^{\mathrm{r}}\tilde l\cos \varpi _{\mathrm{B}}^{\mathrm{r}},m \in \mathcal{M}$ and ${{{{\tilde g}}}_n} = - N_n^{\mathrm{c}}l\cos {\varpi _{\mathrm{c}}} - N_n^{\mathrm{r}}l\cos {\varpi _{\mathrm{r}}},n \in \mathcal{N}$.
\end{Proposition}
{\it{Proof:}} Please refer to Appendix A. ~$\hfill\blacksquare$

According to Proposition \ref{pro:1}, the channel matrix $\bm{G}$ can be reformulated based on the geometric relationship between BS and IRS, and its phase is decomposed into a constant term, a term dependent on the BS antenna index, and a term dependent on the IRS element index.

Let $\theta _{{n}}^{\mathrm{i}}$, $\phi _{{n}}^{\mathrm{i}}$, $\theta _{{n}}^{\mathrm{r}}$, and $\phi _{{n}}^{\mathrm{i}}$ denote the azimuth and elevation angles of the incident wave as well as the azimuth and elevation angles of the reflected wave relative to the $n$-element of the IRS, respectively. Different from the traditional signal-reflection model, we consider an angle-dependent signal reflection model based on electromagnetic-field modeling that takes into account the effective IRS reflection aperture, as represented in the following proposition.
\begin{Proposition}
	\label{pro:2}
	 The reflection coefficient of the $n$-th IRS element is given by
	 \begin{align}
	 	{{\eta }_{{n}}} = \sqrt {\frac{{4\pi {\bar{l}^4}}}{{{\lambda ^2}}}} {{\alpha }_{{n}}}{{\gamma }_{{n}}}{e^{j{{\psi }_{{n}}}}},
	 \end{align}
	 where $\psi _{{n}}$ denotes its phase shift. The reception factor and the reflection factor of the $n$-th IRS element are denoted by ${{\alpha }_{{n}}} = \cos \phi _{{n}}^{\mathrm{i}}$, ${{\gamma }_{{n}}} = {{ Z}_{{n}}}{\operatorname{sinc}} ( {{{X}_{{n}}}} ){\operatorname{sinc}} ( {{{Y}_{{n}}}} )$ with ${{X}_{{n}}} = \frac{{{\pi \bar{l}}}}{\lambda } ( \sin {\theta _{{n}}^{\mathrm{i}}}\cos {\phi _{{n}}^{\mathrm{i}}} + \cos {\theta _{{n}}^{\mathrm{r}}}\sin {\phi _{{n}}^{\mathrm{r}}} )$, ${{ Y}_{{n}}} = \frac{\pi \bar{l}}{\lambda } ( \sin \theta _{{n}}^{\mathrm{r}}\sin \phi _{{n}}^{\mathrm{r}} - \sin \theta _{{n}}^{\mathrm{i}} \sin \phi _{{n}}^{\mathrm{i}} )$, and ${{ Z}_{{n}}} = \sqrt {{{\cos }^2}{\phi _{n}^{\mathrm{r}}}{{\sin }^2} ( {{ \phi _{n}^{\mathrm{i}}} + {\theta _{n}^{\mathrm{r}}}} ) + {{\cos }^2}( {{ \phi _{n}^{\mathrm{i}}} + {\theta _{n}^{\mathrm{r}}}} )} $.
\end{Proposition}

{\it{Proof:}} Please refer to Appendix B. ~$\hfill\blacksquare$

It is observed from Proposition \ref{pro:2} that the reflection coefficient of each IRS element is based on a physical angle-dependent model which characterizes the its inherent electromagnetic response by accounting for the reception factor, ${{\alpha }_{{n}}}$, and the reflection factor, ${{\gamma }_{{n}}}$. Specifically, the reception factor ${{\alpha }_{{n}}}$ indicates the impact of the cosine of the incident azimuth angle on the energy-capturing capability of each element. Moreover, we observe that the reflection factor ${{\gamma }_{{n}}}$ includes two sinc functions, both of which are fundamentally constrained by the element's wavelength-normalized length, i.e., $\bar{l}/\lambda$. 

The received signal power is given by
\begin{align}
	\label{P_r_w}
	{P_{\mathrm{r}}} \!=\! \frac{{4\pi {{\bar l}^4}{\beta ^2}}}{{{\lambda ^2}}}{\left| {\sum\limits_{m = 1}^M {{e^{ - j\frac{{2\pi }}{\lambda }{{{{\bar g}}}_m}}}{w_m}\sum\limits_{n = 1}^N {\frac{{{\alpha _n}{\gamma _n}}}{{{t_{n,m}}{r_n}}}{e^{j( {{\varphi _n} \!-\! \frac{{2\pi }}{\lambda }( {{{{{\tilde g}}}_n} \!+\! {r_n}} )} )}}} } } \right|^2}.
\end{align}
As such, the received SNR at the location $\bm{p}_\mathrm{U} \in \mathcal{A}$ is given by $\varsigma = {{P_{\mathrm{r}}}}/{\sigma _0^2}$. 

\begin{figure}[t]
	\centering
	\includegraphics[width=0.45\textwidth]{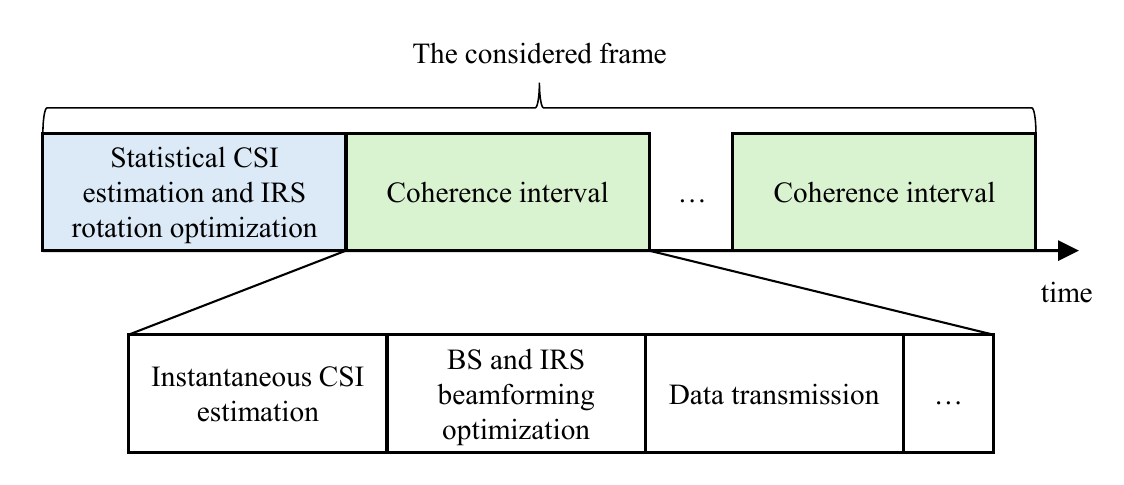}
	\vspace{-5pt}
	\caption{Illustration of the two-timescale design.}
	\label{fig:twotimescale}
	\vspace{-10pt}
\end{figure}
As shown in Fig. \ref{fig:twotimescale}, we propose a two-timescale design, where the short-term transmit beamforming at the BS $\bm{w}$, the short-term IRS reflection phase shifts $\bm{\Phi}$, and the long-term IRS rotation angles $\bm{\Omega} = (\theta, \phi)^T$ are jointly optimized to maximize the worst-case/minimum SNR within the target $\mathcal{A}$. Specifically, the IRS rotation angles are first optimized based on the statistical channel state information (CSI), which depends on the angle and location information that typically remain unchanged over the long term. This is because the IRS rotation is realized through mechanical control, which varies over a large timescale. In contrast, the phase shifts of the IRS are adaptively adjusted for the typical user within the target area, since they can be updated rapidly through electronic control in response to user mobility. Then, the beamforming vectors at both the BS and the IRS are designed based on the effective instantaneous CSI with the optimized IRS rotation obtained in the previous step. Our two-timescale design effectively reduces the update frequency of IRS rotation and allows the system to exploit fast beamforming reconfiguration and robust orientation, thereby ensuring reliable coverage and enhanced signal strength over the target region.

Given $\bm{\Phi}$ and $\bm{\Omega}$, the optimal transmit beamforming at the BS is given by \cite{lu_distance}
\begin{align}
	\label{w}
	{w}_m^{{\mathrm{opt}}} = \sqrt {\frac{{{P_{\mathrm{t}}}}}{M}} {e^{j\frac{{2\pi }}{\lambda }{{{{\bar g}}}_m}}},\forall m \in \mathcal{M}.
\end{align}
It is observed from \eqref{w} that the optimal transmit beamforming at the BS is independent of the user location ${{\bm{p}}_{\mathrm{U}}}$. Given $\{{w}_m^{{\mathrm{opt}}}\}$ and $\bm{\Omega}$, we configure the reflection coefficient matrix of the IRS as
\begin{align}
	{\psi _n^{{\mathrm{opt}}}} = \bmod \left( {\frac{{2\pi }}{\lambda }\left( {{{{{\tilde g}}}_n} + {r_n}} \right),2\pi } \right),\forall n \in \mathcal{N}.
\end{align}
With ${w}_m^{{\mathrm{opt}}},\forall m=1,\cdots, M$ and ${\psi _n^{{\mathrm{opt}}}},\forall n=1,\cdots, N$, the received power is given by ${P_{\mathrm{r}}} = \frac{{4\pi {{\bar l}^4}{\beta ^2}{P_{\mathrm{t}}}}}{{{\lambda ^2}M}}{\left| {\sum\limits_{n = 1}^N {\frac{{{\alpha _n}{\gamma _n}}}{{{r_n}}}\sum\limits_{m = 1}^M {\frac{1}{{{t_{n,m}}}}} } } \right|^2}$.
To facilitate the calculation of channel amplitudes, we consider the far-field case and approximate ${t_{{n,m}}} \approx t$, ${r_{{n}}} \approx r$, $\theta _{{n}}^{\mathrm{i}} \approx \theta ^{\mathrm{i}}$, $\theta _{{n}}^{\mathrm{r}} \approx \theta ^{\mathrm{r}}$, $\phi _{{n}}^{\mathrm{i}} \approx \phi ^{\mathrm{i}}$, $\phi _{{n}}^{\mathrm{r}} \approx \phi ^{\mathrm{r}}$, where $\theta _{{n}}^{\mathrm{i}}$, $\theta _{{n}}^{\mathrm{r}}$, $\phi _{{n}}^{\mathrm{i}}$, and $\phi _{{n}}^{\mathrm{r}}$ are the azimuth and elevation angles of the incident angle and the reflected wave of the center of the IRS. Then, it follows that ${{\alpha }_{{n}}} \approx {\alpha }$, ${{\gamma }_{{n}}} \approx {\gamma }$, ${{q}_{{n}}} \approx {q}$. As such, the received power is expressed as
\begin{align}
	\label{power}
	{P_{\mathrm{r}}} = \frac{{4\pi {{\bar{l}}^4}\beta^2 {P_{\mathrm{t}}}{M}{N^2}{\alpha ^2}{\gamma ^2}}}{{{\lambda ^2}{t^2}{r^2}}}.
\end{align}

It is observed from \eqref{power} that the received power is in the order of $\mathcal{O}(\bar l^4)$, increases linearly with the number of the BS antenna and transmit power, and quadratically with the number of the IRS elements. Moreover, the received power is dependent on the reception and reflection factors of the IRS, which are determined by the incident and reflection angles. In the following, we characterize the relationship between the incident/reflection angles and the IRS rotation angles to highlight the important role of the rotation angles design in the system performance.

The three unit vectors of the local CCS at the IRS are expressed in the global CCS as
\begin{align}
	{\bm{e}}_{\mathrm{x}} =& {\left( {\cos {\theta}\sin {\varphi}, - \sin {\theta _2}\sin {\varphi}, - \cos {\varphi}} \right)^T},\\
	{\bm{e}}_{\mathrm{y}} =& {\left( {\sin {\theta},\cos {\theta},0} \right)^T},\\
	{\bm{e}}_{\mathrm{z}} =& {\left( {\cos {\theta}\cos {\varphi}, - \sin {\theta}\cos {\varphi},\sin {\varphi}} \right)^T}.
\end{align}
Then, the orientation matrix that characterizes the relationship from the global CCS to the local CCS at the IRS is given by
\begin{align}
	\label{CCS}
	{{\bm{Q}}} ( {{\theta},{\varphi}} ) &= ( {{{\bm{e}}_{\mathrm{x}}},{{\bm{e}}_{\mathrm{y}}},{{\bm{e}}_{\mathrm{z}}}} ) \nonumber\\
	&= \left[ {\begin{array}{*{20}{c}}
			{\cos {\theta}\sin {\varphi}}&{\sin {\theta}}&{\cos {\theta}\cos {\varphi}}\\ 
			{ - \sin {\theta}\sin {\varphi}}&{\cos {\theta}}&{ - \sin {\theta}\cos {\varphi}}\\ 
			{ - \cos {\varphi}}&0&{\sin {\varphi}}
	\end{array}} \right] .
\end{align}
In the initial stage, the unit vectors along the $x$-, $y$-, $z$-axis of the local CCS are given by ${\bm{e}}_{\mathrm{x}} = {\left( 0, 0, -1 \right)^T}$, ${\bm{e}}_{\mathrm{y}} = {\left( {0,1,0} \right)^T}$, ${\bm{e}}_{\mathrm{z}} = {\left( {1,0,0} \right)^T}$, respectively, which are aligned with the negative $z$-, $y$-, and $x$-axis unit vectors of the global CCS. Based on \eqref{CCS}, the coordinates in the IRS local CCS that correspond to any 3D location $\bm{p}$ in the global CCS can be expressed as
\begin{align}
	\label{p_L}
	{{\bm{p}}^{{\mathrm{L}}}} = {\bm{Q}}^T\left( {{\theta},{\varphi}} \right)\left( {{\bm{p}} - {{\bm{p}}_\mathrm{c}}} \right) = (x^{\mathrm{L}}, y^{\mathrm{L}}, z^{\mathrm{L}})^T,
\end{align}
where we set ${{\bm{p}_\mathrm{c}}}$ as the IRS reference element at the origin of the corresponding local CCS. The coordinates of the BS and the user in the local CCS at the IRS are given by
\begin{align}
	{\bm{p}}_{\mathrm{B}}^{{\mathrm{L}}} =& {\bm{Q}}^T\left( {{\theta},{\varphi}} \right)\left( {{{\bm{p}}_{\mathrm{B}}} - {{\bm{p}}_{{\mathrm{c}}}}} \right) = {\left[ {x_{\mathrm{B}}^{{\mathrm{L}}},y_{\mathrm{B}}^{{\mathrm{L}}},z_{\mathrm{B}}^{{\mathrm{L}}}} \right]^T}, \label{p_B^L}\\
	{\bm{p}}_{\mathrm{U}}^{{\mathrm{L}}} =& {\bm{Q}}^T\left( {{\theta},{\varphi}} \right)\left( {{{\bm{p}}_{\mathrm{U}}} - {{\bm{p}}_{{\mathrm{c}}}}} \right) = {\left[ {x_{\mathrm{U}}^{{\mathrm{L}}},y_{\mathrm{U}}^{{\mathrm{L}}},z_{\mathrm{U}}^{{\mathrm{L}}}} \right]^T} \label{p_U^L}.
\end{align}
Therefore, the azimuth angle of the incident angle ${{\theta }^{\mathrm{i}}} \in \left[0,2\pi\right]$, the azimuth angle of the reflected wave of the center of the IRS ${{\theta }^{\mathrm{r}}} \in \left[0,2\pi\right]$, the elevation angle of the incident angle ${{\phi }^{\mathrm{i}}} \in \left[0,\pi/2\right]$, and the elevation angle of the reflected wave of the center of the IRS ${{\phi }^{\mathrm{r}}} \in \left[0,\pi/2\right]$ are rewritten as
\begin{align}
	{{\theta }^{\mathrm{i}}} =& \arctan\frac{{- y_{\mathrm{B}}^{{\mathrm{L}}}}}{{x_{\mathrm{B}}^{{\mathrm{L}}}}}, \label{theta_i_1}\\
	{{\theta }^{\mathrm{r}}} =& \arctan\frac{{y_{\mathrm{U}}^{{\mathrm{L}}}}}{{x_{\mathrm{U}}^{{\mathrm{L}}}}},\label{theta_r_1}\\
	{{\phi }^{\mathrm{i}}} =& \arccos \frac{{z_{\mathrm{B}}^{{\mathrm{L}}}}}{{\left\| {{{\bm{p}}_{\mathrm{B}}} - {{\bm{p}}_{{\mathrm{c}}}}} \right\|}},\label{phi_i_1}\\
	{{\phi }^{\mathrm{r}}} =& \arccos \frac{{z_{\mathrm{U}}^{{\mathrm{L}}}}}{{\left\| {{{\bm{p}}_{{\mathrm{U}}}} - {{\bm{p}}_{{\mathrm{c}}}}} \right\|}}\label{phi_r_1}.
\end{align}
Based on \eqref{p_B^L} and \eqref{p_U^L}, \eqref{theta_i_1}, we can express \eqref{theta_r_1}, \eqref{phi_i_1}, and \eqref{phi_r_1} with respect to $\theta$ and $\phi$ as
\begin{align}
	{{\theta }^{\mathrm{i}}} \!\!=&\! \arctan \! \frac{{- {x_{\mathrm{B}}}\sin \theta - ( {{y_{\mathrm{B}}} - {y_{\mathrm{c}}}} )\cos \theta }}{{{x_{\mathrm{B}}}\cos \theta \sin \varphi \!\!-\!\! ( {{y_{\mathrm{B}}} \!\!-\!\! {y_{\mathrm{c}}}} )\! \sin \theta \sin \varphi \!\!+\!\! {z_{\mathrm{c}}}\cos \varphi }}, \label{theta_i}\\
	{{\theta }^{\mathrm{r}}} \!\!=&\! \arctan \! \frac{{{x_{\mathrm{U}}}\sin \theta +( {{y_{\mathrm{U}}} - {y_{\mathrm{c}}}} )\cos \theta }}{{{x_{\mathrm{U}}}\cos \theta \sin \varphi \!\!-\!\! ( {{y_{\mathrm{U}}} \!\!-\!\! {y_{\mathrm{c}}}} )\sin \theta \sin \varphi \!\!+\!\! {z_{\mathrm{c}}}\cos \varphi }}, \label{theta_r}\\
	{{\phi }^{\mathrm{i}}} \!\!=&\! \arccos \! \frac{{{x_{\mathrm{B}}}\cos \theta \cos \varphi \!\!-\!\! ( {{y_{\mathrm{B}}} \!\!-\!\! {y_{\mathrm{c}}}} )\sin \theta \cos \varphi \!\!-\!\! {z_{\mathrm{c}}}\sin \varphi }}{{\sqrt {x_{\mathrm{B}}^2 + {{( {{y_{\mathrm{B}}} - {y_{\mathrm{c}}}} )}^2} + z_{\mathrm{c}}^2} }}, \label{phi_i}\\
	{{\phi }^{\mathrm{r}}} \!\!=&\! \arccos \! \frac{{{x_{\mathrm{U}}}\cos \theta \cos \varphi \!\!-\!\! \sin \theta \cos \varphi ( {{y_{\mathrm{U}}} \!\!-\!\! {y_{\mathrm{c}}}} ) \!\!-\!\! {z_{\mathrm{c}}}\sin \varphi }}{{\sqrt {x_{\mathrm{U}}^2 + {{( {{y_{\mathrm{U}}} - {y_{\mathrm{c}}}} )}^2} + z_{\mathrm{c}}^2} }}. \label{phi_r}
\end{align}

We denote the normalized incident vector from the BS to the IRS and the normalized reflection vector from the IRS to the user as $\bm{ a}_{\mathrm{t}}$ and $\bm{ a}_{\mathrm{r}}$, respectively, which are given by 
\begin{align}
	&\bm{ a}_{\mathrm{t}} = \frac{{{{\bm{p}}_{{\mathrm{c}}}} - {{\bm{p}}_{\mathrm{B}}}}}{{\left\| {{{\bm{p}}_{{\mathrm{c}}}} - {{\bm{p}}_{\mathrm{B}}}} \right\|}}, \\
	&{{{\bm{a}}}_{\mathrm{r}}}= \frac{{{{\bm{p}}_{{\mathrm{U}}}} - {{\bm{p}}_{{\mathrm{c}}}}}}{{\left\| {{{\bm{p}}_{{\mathrm{U}}}} - {{\bm{p}}_{{\mathrm{c}}}}} \right\|}}.
\end{align}
To enable effective reflection, the BS and the user must be located in front of the IRS, which yields
\begin{align}
	{{\bm{a}}_{\mathrm{t}}^T{{\bm{k}}}} \ge 0, \;\; {{\bm{ a}}_{\mathrm{r}}^T{{\bm{k}}}} \le 0.
\end{align}
Based on \eqref{p_L}, we have $z_{\mathrm{B}}^{\mathrm{L}} \ge 0$ and $z_{\mathrm{U}}^{\mathrm{L}} \ge 0$. Then, it follows that
\begin{align}
	&( {{x_{\mathrm{B}}}\cos \theta \cos \varphi \!-\! ( {{y_{\mathrm{B}}} \!-\! {y_{\mathrm{c}}}} )\sin \theta \cos \varphi \!-\! {z_{\mathrm{c}}}\sin \varphi } ) \ge 0, \nonumber\\
	&( {{x_{\mathrm{U}}}\cos \theta \cos \varphi \!-\! ( {{y_{\mathrm{U}}} \!-\! {y_{\mathrm{c}}}} ) \sin \theta \cos \varphi \!-\! {z_{\mathrm{c}}}\sin \varphi } ) \ge 0. \label{reflection}
\end{align}

Given $\{w_m^\mathrm{opt}\}$ and $\{\phi_n^\mathrm{opt}\}$, our objective is to maximize the minimum SNR over all locations within the target area $\mathcal{A}$ by optimizing the rotation angles of the IRS. Therefore, the optimization problem can be formulated as
\begin{align}
	\label{pro:power}
	\mathop {\max }\limits_{\bm{\Omega} \in \mathcal{S}} \mathop {\min }\limits_{\bm{p}_\mathrm{U} \in \mathcal{A}} \; \varsigma \;\;\;\;\;\; \mathrm{s.t.} \; \eqref{reflection}.
\end{align}
Problem \eqref{pro:power} is intractable for two main reasons. First, the objective function is defined as the minimum received power over a continuous region, which yields a semi-infinite and non-smooth formulation and generally lacks a closed-form expression of the IRS rotation angles. Second, problem \eqref{pro:power} is non-convex, since the objective function is non-concave and the constraints are non-convex.

\section{Single Target Location SNR Maximization}
\label{Single Target Location SNR Optimization}
In this section, we consider the special case of single target location power enhancement, where the target area $\mathcal{A}$ reduces to the point $\bm{p}_\mathrm{s}$. In this case, the inner minimization within problem \eqref{pro:power} can be omitted. Thus, problem \eqref{pro:power} can be reformulated as
\begin{align}
		\label{pro:power_single}
		\mathop {\max }\limits_{\bm{\Omega} \in \mathcal{S}} \; \varsigma \;\;\;\;\;\;
		\mathrm{s.t.} \; \eqref{reflection}.
\end{align}
Let $\bar{L} = \bar l/\lambda $ denote the wavelength-normalized length of each IRS element. With \eqref{power}, problem \eqref{pro:power_single} is reduced to
\begin{align}
		\label{pro:power_1}
		\mathop {\max }\limits_{\bm{\Omega} \in \mathcal{S}} \; \delta_1 \delta_2 \;\;\;\;\;\;
		\mathrm{s.t.} \; \eqref{reflection},
\end{align}
where $\delta _1 = {\cos ^2}{\phi ^{\mathrm{i}}} \left( {{{\cos }^2}{\phi ^{\mathrm{r}}}{{\cos }^2} \left( {{\theta ^{\mathrm{i}}} + {\theta ^{\mathrm{r}}}} \right) + {{\sin }^2} \left( {{\theta ^{\mathrm{i}}} + {\theta ^{\mathrm{r}}}} \right)} \right)$ and $\delta _2 = \operatorname{sinc}^2 \left( \pi \bar{L} \left( {\cos {{ \theta }^{\mathrm{i}}}\sin {{ \phi }^{\mathrm{i}}} + \sin {{ \theta }^{\mathrm{r}}}\cos {{ \phi }^{\mathrm{r}}}} \right) \right) \times \operatorname{sinc}^2 \left(\pi \bar{L} \left( \sin {{ \theta }^{\mathrm{r}}}\sin {{ \phi }^{\mathrm{r}}} - \sin {{ \theta }^{\mathrm{i}}}\sin {{ \phi }^i} \right) \right)$.

\begin{figure*}[!t]
	\centering
	\subfloat[$\delta_1$]{
		\includegraphics[width=0.25\textwidth]{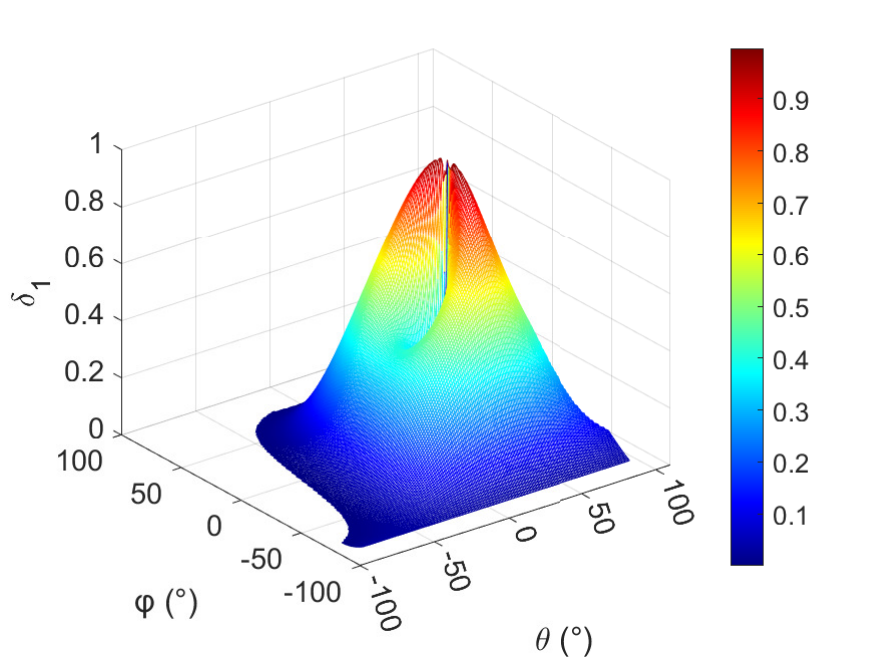}}
	\subfloat[$\delta_2$]{
		\includegraphics[width=0.25\textwidth]{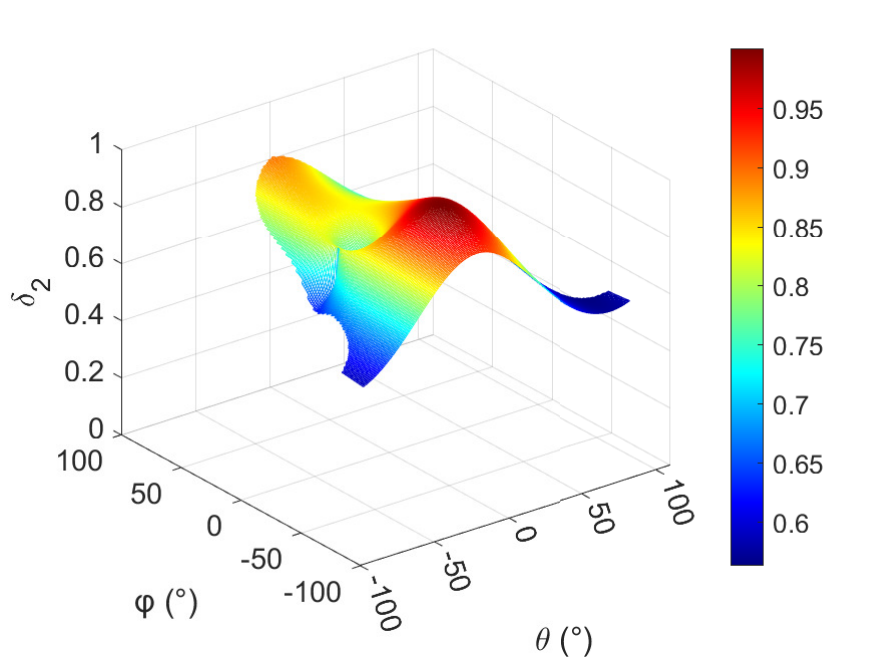}}
	\subfloat[$\delta_1 \times \delta_2$]{
		\includegraphics[width=0.25\textwidth]{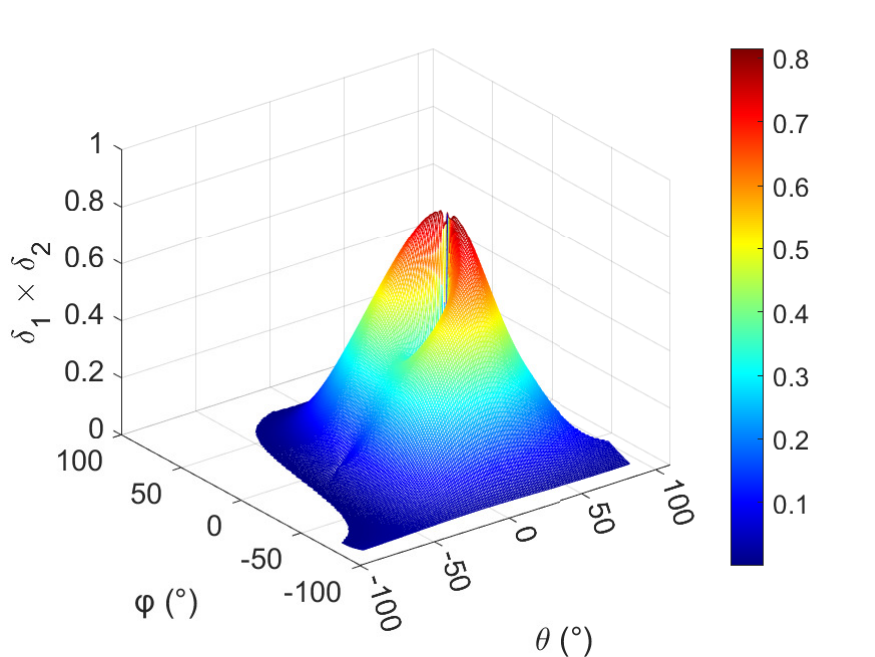}}
	\caption{Values of $\delta_1$, $\delta_2$, and $\delta_1 \times \delta_2$ versus the rotation $\bm{\Omega}$ when $\bar{L} = 0.25$.}
	\label{fig:0.25}
\end{figure*}
\begin{figure*}[!t]
	\vspace{-20pt}
	\centering
	\subfloat[$\delta_1$]{
		\includegraphics[width=0.25\textwidth]{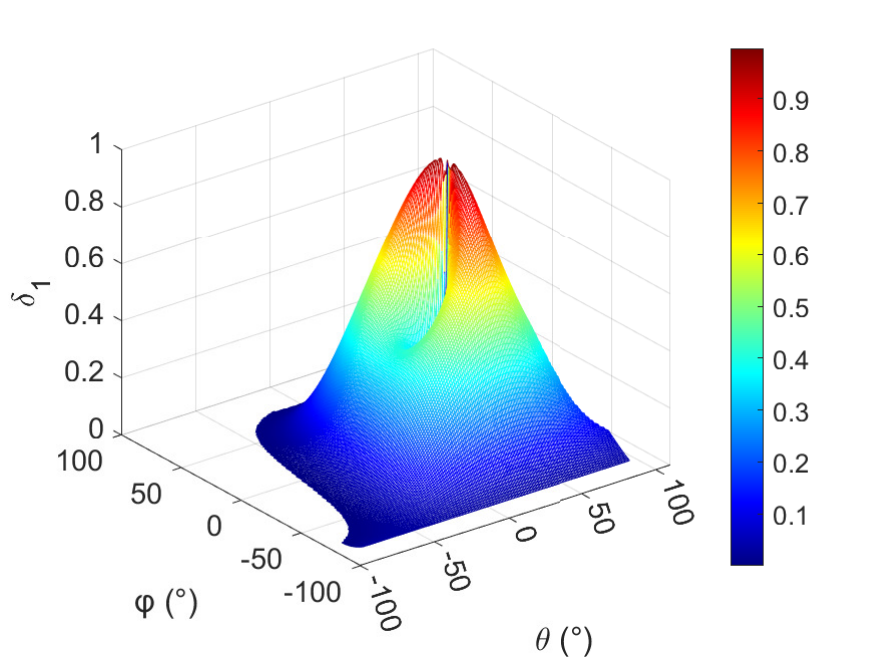}}
	\subfloat[$\delta_2$]{
		\includegraphics[width=0.25\textwidth]{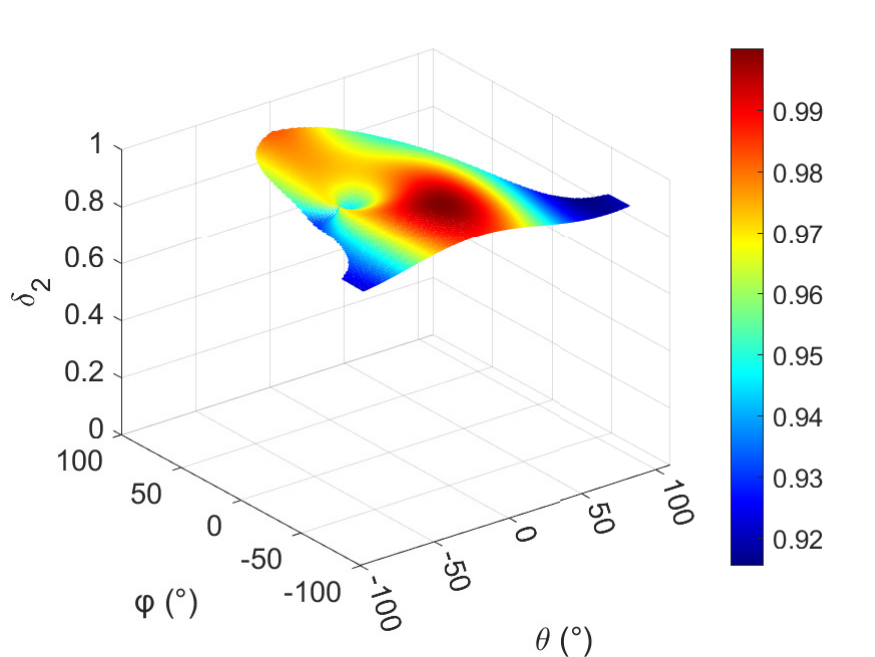}}
	\subfloat[$\delta_1 \times \delta_2$]{
		\includegraphics[width=0.25\textwidth]{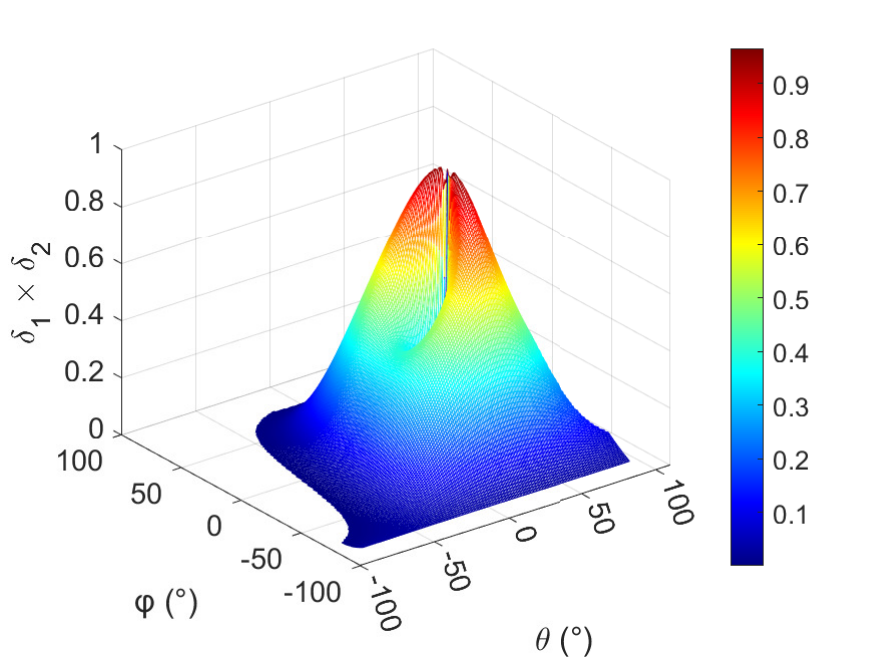}}
	\caption{Values of $\delta_1$, $\delta_2$, and $\delta_1 \times \delta_2$ versus the rotation $\bm{\Omega}$ when $\bar{L} = 0.1$.}
	\label{fig:0.5}
\end{figure*}

The objective function of problem \eqref{pro:power_1} consists of two terms $\delta_1$ and $\delta_2$, where $\delta_2$ composed of the $\operatorname{sinc}(\cdot)$ function with the wavelength-normalized length $\bar L$. Without loss of generality, the IRS consists of a large number of passive reflecting elements with sub-wavelength inter-element spacing, i.e., $l = 0.5\lambda > \bar{l}$. Then, we have $\bar {L} < 0.5$. Motivated by the presence of the term $\delta_2$ and the fact that $\operatorname{sinc}(\cdot) \approx 1$ for small arguments, we characterize the impact of $\bar{L}$ on the objective function. Figs. \ref{fig:0.25} and \ref{fig:0.5} illustrate the values of $\delta_1$, $\delta_2$, and $\delta_1\delta_2$ versus the rotation angles $\bm{\Omega}$ when $\bar{L} = 0.25$ and $\bar{L} = 0.1$, respectively. The BS, IRS, and user are located at $\bm{p}_\mathrm{B} = [50,20,0]^T$, $\bm{p}_\mathrm{c} = [0,50,10]^T$, and $\bm{p}_\mathrm{s} = [30,80,0]^T$, respectively. It is observed from Figs. \eqref{fig:0.25} and \eqref{fig:0.5} that $\delta_2$ exhibits less significant variations as the rotation angles change, as compared to $\delta_1$. This motivates us to focus on the rotation angles $\bm{\Omega}$ that are designed to maximize $\delta_1$ in the following. Problem \eqref{pro:power_1} can be approximated as
\begin{subequations}
	\label{pro:power_small}
	\begin{align}
		\mathop {\max }\limits_{\bm{\Omega} \in \mathcal{S}} \;\;\;& {\cos ^2}{\phi ^{\mathrm{i}}}\left( {{{\cos }^2}{\phi ^{\mathrm{r}}}{{\cos }^2}\left( {{\theta ^{\mathrm{i}}} + {\theta ^{\mathrm{r}}}} \right) + {{\sin }^2}\left( {{\theta ^{\mathrm{i}}} + {\theta ^{\mathrm{r}}}} \right)} \right) \\
		\mathrm{s.t.} \;\;\;
		&\eqref{reflection}. 
	\end{align}
\end{subequations}

Problem \eqref{pro:power_small} is challenging to solve directly, since the objective function involves the complicated expression of the incident and reflection angles, which are functions of the optimization variables $\bm{\Omega} = (\theta, \phi)^T$. We propose a tractable approach that yields a near-optimal solution, as expressed in the following proposition.
\begin{Proposition}
	\label{small}
	Under the assumption that $\frac{{\left( {{x_{\mathrm{B}}}\left( {{y_{\mathrm{U}}} - {y_{\mathrm{c}}}} \right) - {x_{\mathrm{U}}}\left( {{y_{\mathrm{B}}} - {y_{\mathrm{c}}}} \right)} \right)\sqrt {x_{\mathrm{B}}^2 + {{\left( {{y_{\mathrm{B}}} - {y_{\mathrm{c}}}} \right)}^2} + z_{\mathrm{c}}^2} }}{{{z_{\mathrm{c}}}\left( {x_{\mathrm{B}}^2 + {{\left( {{y_{\mathrm{B}}} - {y_{\mathrm{c}}}} \right)}^2} - {x_{\mathrm{B}}}{x_{\mathrm{U}}} - \left( {{y_{\mathrm{U}}} - {y_{\mathrm{c}}}} \right)\left( {{y_{\mathrm{B}}} - {y_{\mathrm{c}}}} \right)} \right)}} \gg 1$, the optimized azimuth angle and elevation angle of the IRS can be given by $\theta^{\mathrm{opt}} = \arctan \left( { - \frac{{{y_{\mathrm{B}}} - {y_{\mathrm{c}}}}}{{{x_{\mathrm{B}}}}}} \right)$ and $\varphi^{\mathrm{opt}} = \arctan \left( {\frac{{ - {z_{\mathrm{c}}}}}{{\sqrt {x_{\mathrm{B}}^2 + {{( {{y_{\mathrm{B}}} - {y_{\mathrm{c}}}})^2}}} }}} \right)$.
\end{Proposition}
{\it{Proof:}} Please refer to Appendix C.
~$\hfill\blacksquare$

It is observed from Proposition \ref{small} that the assumption is readily satisfied, especially when $z_\mathrm{c}$ approaches 0. In this case, the design of $\theta$ determines the system performance without the need for elevation rotation, i.e., $\varphi^\mathrm{opt} = 0$. Moreover, the optimized elevation angle decreases as $z_\mathrm{c}$ increases while the azimuth angle remains unchanged. If the altitude of the IRS is sufficiently large, i.e., $z_\mathrm{c} \to \infty$, we have ${\phi ^{\mathrm{i}}} = {\phi ^{\mathrm{r}}} \to \arccos \left(-\sin \varphi\right)$. The optimized azimuth angle of the IRS is approximated as $\varphi^{\mathrm{opt}} = -\pi/2$ regardless of the value of $\theta$ while ensuring that constraint \eqref{reflection} holds. This is expected because both BS and the user are distributed on the ground, which is far from the IRS. The IRS acts as a point source, and beamforming gains depend primarily on optimizing $\phi$ rather than $\theta$. If both BS and user are located far along the $x$-axis from the IRS, i.e., $x_\mathrm{B} \to \infty$ and $x_\mathrm{U} \to \infty$, it follows that ${\phi ^{\mathrm{i}}} = {\phi ^{\mathrm{r}}} \to \arccos \left(\cos \theta \cos \varphi\right)$. The optimal azimuth and elevation angles of the IRS are approximated as $\theta^{\mathrm{opt}} = \varphi^{\mathrm{opt}} = 0$. This implies that no rotation is needed, for similar reasons as in the case of $z_\mathrm{c} \to \infty$.

By exploiting the specific structures of problem \eqref{pro:power_1} and guided by the insights in Proposition \ref{small}, we propose an efficient PSO-based method, due to the following main issues: First, the particle swarm evaluates multiple candidates in parallel and cooperatively concentrates in the neighborhood of the near-optimal solution. Second, given the rotation constraints that ensures effective reflection, physical feasibility can be incorporated into PSO by masking or projecting particles, thus avoiding wasted evaluations in infeasible sectors. Third, in the continuous angle domain rather than via grid search, PSO preserves global exploration and supports local refinement, which yields both robust global solutions and high-precision angular resolution. The processes of the proposed algorithm are provided as follows. The swarm is initialized with a high proportion of particles sampled in a small neighborhood around the high-quality solution, with the remaining particles sampled uniformly over the feasible domain. Thus, high-probability regions are reliably identified, and faster convergence is achieved than conventional PSO with random initialization. To ensure the effective reflection, i.e., constraint \eqref{reflection}, each particle's performance is determined by a fitness function with an introduced penalty term as
\begin{align}
	\label{fitness}
	\mathcal{L}\left(\bm{\Omega}\right) = \delta_1\delta_2 - \tau (\max \left\{0, -z_\mathrm{B}^\mathrm{L}\right\} + \max \left\{0, -z_\mathrm{U}^\mathrm{L}\right\}),
\end{align}
where $\tau$ is a penalty parameter satisfying $\tau > 0$. Let $\bm{\Omega}_b^{\left(t\right)}$ and $\bm{\nu}_b^{\left(t\right)}$ denote the position and velocity of the $b$-th particle at the $t$-th iteration. In each iteration, the position of each particle is updated as
\begin{align}
	\label{position}
	\bm{\Omega}_b^{\left(t+1\right)} = \mathcal{G}\left(\bm{\Omega}_b^{\left(t\right)} + \bm{\nu}_b^{\left(t\right)}\right),
\end{align}
where $\mathcal{G}\left(\cdot\right)$ is an element-wise projection operator that ensures $\bm{\Omega}_b^{\left(t+1\right)} \in \mathcal{S}$. Let $\bm{\Omega}_{b,\mathrm{lbest}}$ and $\bm{\Omega}_\mathrm{gbest}$ denote the local best position of the $b$-th particle and the global best position among all particles, respectively. In the $(t+1)$-th iteration, the velocity of the $b$-th particle can be expressed as
\begin{align}
	\label{velocity}
	\bm{\nu}_b^{\left(t+1\right)} &= \omega^{\left(t\right)}\bm{\nu}_b^{\left(t\right)} + c_1 \cdot \operatorname{rand} \cdot (\bm{\Omega}_{b,\mathrm{lbest}} - \bm{\Omega}_b^{\left(t\right)}) \nonumber\\
	&+ c_2 \cdot \operatorname{rand} \cdot (\bm{\Omega}_{\mathrm{gbest}} - \bm{\Omega}_b^{\left(t\right)}),
\end{align}
where $c_1$ and $c_2$ denote the cognitive and social coefficients, respectively. To balance the accuracy and convergence speed of the PSO, the inertia weight is denoted by
\begin{align}
	\omega^{\left(t\right)} = \left(\omega_\mathrm{ini} - \omega_\mathrm{end}\right) {(T_\mathrm{max}-t)}/{T_\mathrm{max}} + \omega_\mathrm{end},
\end{align}
where $T_\mathrm{max}$ is the number of iterations, $\omega_\mathrm{ini}$ is the initial inertia weight, and $\omega_\mathrm{end}$ is the final inertia weight. Based on both its own and the global best position, each particle iteratively updates its personal best position. Finally, the global best position is then updated by selecting the particle with the highest fitness value, ensuring a non-decreasing fitness value throughout the optimization process until convergence. The main procedures of the proposed PSO-based algorithm are summarized in Algorithm \ref{Algo:single} with the computational complexity of $\mathcal{O}\left(2BT_{\mathrm{max}}\right)$.

\begin{algorithm}[t]
	\small
	\label{Algo:single}
	\SetAlgoLined 
	\caption{PSO-based Algorithm for Solving Problem \eqref{pro:power_1}}
	\KwIn{Number of particles $B$, maximum number of iterations $T_\mathrm{max}$, the cognitive coefficients $c_1$, the social coefficients $c_2$, the initial inertia weight $\omega_\mathrm{ini}$, the final inertia weight $\omega_\mathrm{end}$, $\bm{p}_\mathrm{B}$, $\bm{p}_\mathrm{c}$, $\bm{p}_\mathrm{U}$, and $\bar L$.}
	\KwOut{$\bm{\Omega}^\mathrm{opt}$.}
	Initialize particle positions $\{\bm{\Omega}_1^{\left(0\right)},\cdots,\bm{\Omega}_B^{\left(0\right)}\}$, and velocities $\{\bm{\nu}_1^{\left(0\right)},\cdots,\bm{\nu}_B^{\left(0\right)}\}$, local best positions $\bm{\Omega}_{b,\mathrm{lbest}}$, and global best position $\bm{\Omega}_\mathrm{gbest}$.\\
	\For{$t = 1$ to $T_\mathrm{max}$}{
		\For{$b = 1$ to $B$}{
			Update $\bm{\Omega}_b^{\left(t\right)}$ via \eqref{position} and $\bm{\nu}_b^{\left(t\right)}$ via \eqref{velocity}.
			
			\If{$\mathcal{L} (\bm{\Omega}_b^{\left(t\right)}) > \mathcal{L} \left(\bm{\Omega}_{b,\mathrm{lbest}}\right)$}{
				$\bm{\Omega}_{b,\mathrm{lbest}} \leftarrow \bm{\Omega}_b^{\left(t\right)}$
			}
			
			\If{$\mathcal{L} \left(\bm{\Omega}_{b,\mathrm{lbest}}\right) > \mathcal{L} \left(\bm{\Omega}_{\mathrm{gbest}}\right)$}{
				$\bm{\Omega}_{\mathrm{gbest}} \leftarrow \bm{\Omega}_{b,\mathrm{lbest}}$
			}
		}
	}
	Return $\bm{\Omega}^\mathrm{opt} = \bm{\Omega}_{\mathrm{gbest}}$.
\end{algorithm}

\section{Area Coverage Enhancement}
\label{Area Coverage Enhancement}
In this section, we consider the rotatable IRS-aided area coverage enhancement. Based on \eqref{power}, problem \eqref{pro:power} can be reduced to
\begin{align}
	\label{pro:power_area}
	\mathop {\max }\limits_{\bm{\Omega} \in \mathcal{S}} \mathop {\min }\limits_{\bm{p}_\mathrm{U} \in \mathcal{A}} \; \frac{{\delta _1}{\delta _2}}{\| {{{\bm{p}}_{\mathrm{U}}} - {{\bm{p}}_{\mathrm{c}}}} \|^2} \;\;\;\; 
	\mathrm{s.t.} \; \eqref{reflection}.
\end{align}	
Problem \eqref{pro:power_area} is still challenging to solve optimally via the conventional optimization methods. To solve this problem, we propose a PSO-based two-loop iterative algorithm in the following. 

In the inner loop of our proposed algorithm, we calculate the value of the fitness function for each particle, which is given by 
\begin{align}
	\mathcal{D}(\bm{\Omega}) \!\!=\!\! \frac{{\delta _1}{\delta _2}}{\| {{{\bm{p}}_{\mathrm{U}}} \!-\! {{\bm{p}}_{\mathrm{c}}}} \|^2} \!-\! \tau (\max \{0,\! -z_\mathrm{B}^\mathrm{L}\} \!+\! \max \{0,\! -z_\mathrm{U}^\mathrm{L}\}).
\end{align}
Thus, it follows that
\begin{align}
	\label{pro:min_area}
	\mathop {\min }\limits_{\bm{p}_\mathrm{U} \in \mathcal{A}} \; \mathcal{D}({\bm \Omega_{b}^{(t)}}), 1 \le b \le B, 1 \le t \le T_\mathrm{\max}.
\end{align}
By exploiting the specific structure of problem \eqref{pro:min_area}, we first obtain the null points of $\delta_2$. For any rotation $\bm{\Omega}$, let $\Delta_1^{\max}$ and $\Delta_2^{\max}$ denote the maximum projection values, as well as $\Delta_1^{\min}$ and $\Delta_2^{\min}$ denote the minimum projection values within the target area $\mathcal{A}$, which are given by
\begin{align}
	\Delta_1^{\max} &= \mathop {\max }\limits_{\bm{p}_\mathrm{U} \in \mathcal{A}} {\cos {\theta ^{\mathrm{i}}}\sin {\phi ^{\mathrm{i}}} + \sin {\theta ^{\mathrm{r}}}\cos {\phi ^{\mathrm{r}}}}, \\
	\Delta_2^{\max} &= \mathop {\max }\limits_{\bm{p}_\mathrm{U} \in \mathcal{A}} {\sin {\theta ^{\mathrm{r}}}\sin {\phi ^{\mathrm{r}}} - \sin {\theta ^{\mathrm{i}}}\sin {\phi ^i}}, \\
	\Delta_1^{\min} &= \mathop {\min }\limits_{\bm{p}_\mathrm{U} \in \mathcal{A}} {\cos {\theta ^{\mathrm{i}}}\sin {\phi ^{\mathrm{i}}} + \sin {\theta ^{\mathrm{r}}}\cos {\phi ^{\mathrm{r}}}}, \\
	\Delta_2^{\min} &= \mathop {\min }\limits_{\bm{p}_\mathrm{U} \in \mathcal{A}} {\sin {\theta ^{\mathrm{r}}}\sin {\phi ^{\mathrm{r}}} - \sin {\theta ^{\mathrm{i}}}\sin {\phi ^i}}.
\end{align}
The null points of $\delta_2$ can be derived by setting $\pi \bar{L}\Delta_1 = \pm v_1\pi$, $v_1 \in \mathcal{N}_{+}$, i.e., $\Delta_1 = \pm v_1/\bar{L}$, or $\pi \bar{L}\Delta_2 = \pm v_2\pi$, $v_2 \in \mathcal{N}_{+}$, i.e., $\Delta_2 = \pm v_2/\bar{L}$. For any rotation $\bm{\Omega}$, the minimum value of $\delta_2$ within $\mathcal{A}$ is zero if there exist null points. It implies that the minimum value of $\delta_1 \delta_2$ within $\mathcal{A}$ is zero. By prioritizing the detection of null points within the target region, i.e., $\delta_2 = 0$, it avoids exhaustive grid search over all user positions, thereby significantly reducing the computational complexity. 

In the outer loop of our proposed algorithm, Algorithm \ref{Algo:single} can be applied to optimize the reflection angle, which is designed to iteratively update the particle positions based on their individual and collective experiences. For each particle, the corresponding fitness value is evaluated in the inner loop. 

The main procedures for solving problem \ref{pro:power_area} are summarized in Algorithm \eqref{Algo:area}. The complexity of the inner-loop is $\mathcal{O}(A_\mathrm{x}A_\mathrm{y})$ if no null point is detected. In
the outer-loop, the complexity of the PSO-based method is $\mathcal{O}(2BT_\mathrm{\max})$. Therefore, the overall complexity of our proposed algorithm is $\mathcal{O}(2BT_\mathrm{\max}A_\mathrm{x}A_\mathrm{y})$ in the worst case without null points.

\begin{algorithm}[t]
	\small
	\label{Algo:area}
	\SetAlgoLined
	\caption{PSO-based Two-Loop Iterative Algorithm}
	\KwIn{Number of particles $B$, maximum number of iterations $T_\mathrm{max}$, the cognitive coefficients $c_1$, the social coefficients $c_2$, the initial inertia weight $\omega_\mathrm{ini}$, the final inertia weight $\omega_\mathrm{end}$, $\bm{p}_\mathrm{B}$, $\bm{p}_\mathrm{c}$, $\bm{p}_\mathrm{U}$, $A_\mathrm{x}$, $A_\mathrm{y}$, and $\bar L$.}
	\KwOut{$\bm{\Omega}^\mathrm{opt}$.}
	Initialize particle positions $\{\bm{\Omega}_1^{\left(0\right)},\cdots,\bm{\Omega}_B^{\left(0\right)}\}$, and velocities $\{\bm{\nu}_1^{\left(0\right)},\cdots,\bm{\nu}_B^{\left(0\right)}\}$, local best positions $\bm{\Omega}_{b,\mathrm{lbest}}$, and global best position $\bm{\Omega}_\mathrm{gbest}$.\\
	\For{$t = 1$ \KwTo $T_{\mathrm max}$ }{
		\For{$b = 1$ \KwTo $B$}{
			Update $\bm{\Omega}_b^{\left(t\right)}$ and $\bm{\nu}_b^{\left(t\right)}$.
			
			\For{$x_\mathrm{U} = x_\mathrm{r} - A_\mathrm{x}/2$ \KwTo $x_\mathrm{r} + A_\mathrm{x}/2$}{
				\For{$y_\mathrm{U} = y_\mathrm{r} - A_\mathrm{y}/2$ \KwTo $y_\mathrm{r} + A_\mathrm{y}/2$}{
					Calculate ${{\theta }^{\mathrm{i}}}$, ${{\theta }^{\mathrm{r}}}$, ${{\phi }^{\mathrm{i}}}$, and ${{\phi }^{\mathrm{i}}}$ based on \eqref{theta_i}, \eqref{theta_r}, \eqref{phi_i}, and \eqref{phi_r}.\\
					Calculate $\Delta_1$ and $\Delta_2$.\\
					Update $\Delta_1^{\min}$, $\Delta_1^{\max}$, $\Delta_2^{\min}$, and $\Delta_2^{\max}$.
				}
			}
			Obtain the minimum value of $\delta_1 \delta_2$ within $\mathcal{A}$ as $\mathcal{L} (\bm{\Omega}_b^{\left(t\right)})$.
			
			\If{$\mathcal{D} (\bm{\Omega}_b^{\left(t\right)}) > \mathcal{D} \left(\bm{\Omega}_{b,\mathrm{lbest}}\right)$}{
				$\bm{\Omega}_{b,\mathrm{lbest}} \leftarrow \bm{\Omega}_b^{\left(t\right)}$
			}
			
			\If{$\mathcal{D} \left(\bm{\Omega}_{b,\mathrm{lbest}}\right) > \mathcal{D} \left(\bm{\Omega}_{\mathrm{gbest}}\right)$}{
				$\bm{\Omega}_{\mathrm{gbest}} \leftarrow \bm{\Omega}_{b,\mathrm{lbest}}$
			}
			
		}
	}
	Return $\bm{\Omega}^\mathrm{opt} = \bm{\Omega}_{\mathrm{gbest}}$.
\end{algorithm}

\section{Simulation Results}
\label{Simulation Results}
In this section, simulation results are provided to evaluate the performance of our proposed design. The center of the BS, IRS, and target area, as well as the single target point are located at $\bm{p}_\mathrm{B} = [50,20,0]^T$ m, $\bm{p}_\mathrm{c} = [0,50,10]^T$ m, $\bm{p}_\mathrm{r} = [30,80,0]^T$ m, and $\bm{p}_\mathrm{s} = [30,80,0]^T$ m, respectively. The reference channel power gain is $\beta = -40$ dB. Other system parameters are set as: $M = 128$, $N = 256$, $\bar{L}=0.25$, $A_\mathrm{x} = A_\mathrm{y} = 10$ m, $P_\mathrm{t} = 30$ dBm, and $\sigma_0^2 = -90$ dBm. In the PSO-based algorithm, we set $B = 100$, $T_\mathrm{max} = 30$, $c_1 = c_2 = 2$, $\omega_\mathrm{ini} = 0.4$, $\omega_\mathrm{end} = 0.9$, and $\bm{\nu}_b^{\left(t\right)} \in \left[-5,5\right]$. For comparison, the following schemes are considered: 1) Fixed $\varphi$: the IRS can adjust its azimuth angle, while its elevation angle is fixed at $\varphi = 10^{\circ}$; 2) Fixed $\theta$: the IRS can adjust its elevation angle, while its azimuth angle is fixed at $\theta = 10^{\circ}$; 3) Fixed rotation: the rotation angles of the IRS is fixed at $\bm{\Omega} = \left\{0^{\circ},0^{\circ}\right\}$; 4) M-IRS: the IRS can move along $y$-axis with fixed rotation angles $\bm{\Omega} = \left\{0^{\circ},0^{\circ}\right\}$.


\subsection{Single Target Location SNR Enhancement}
\subsubsection{Evaluation of the Proposed Schemes}

\begin{figure}[t]
	\centering
	\includegraphics[width=0.75\linewidth]{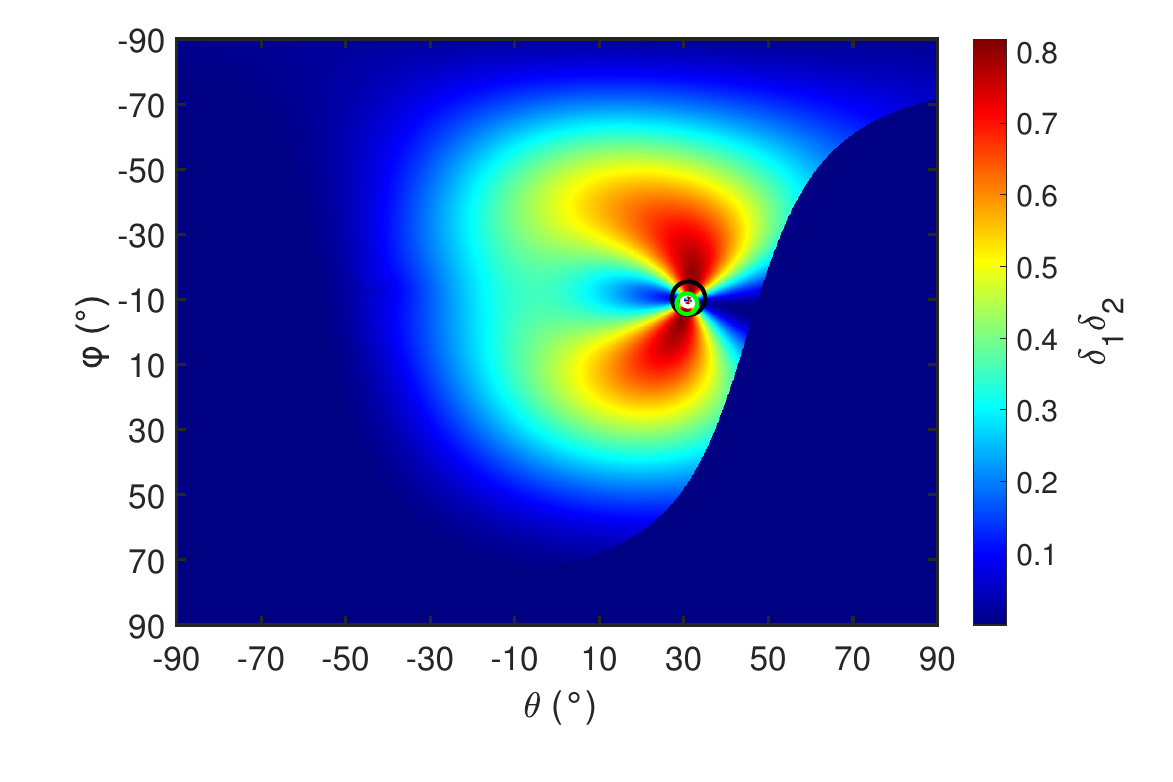}
	\vspace{-15pt}
	\caption{$\delta_1\delta_2$ versus the azimuth angle $\theta$ and elevation angle $\varphi$, where the optimized rotation angles $\bm{\Omega}^\mathrm{opt} = \left\{\theta^\mathrm{opt}, \varphi^\mathrm{opt}\right\}$ obtained by the exhaustive search method, Algorithm \ref{Algo:single}, and the closed form in Proposition \ref{small} are marked with black, green, and white circles, respectively.}
	\label{fig:closedform}
\end{figure}
In Fig. \ref{fig:closedform}, we consider the following schemes: 1) ES: the rotation angles is optimized via the the exhaustive search method; 2) Proposed Opt: the rotation angles is optimized via Algorithm \ref{Algo:single}; 3) Proposed Subopt: the rotation angles is obtained in Proposition \ref{small}. Fig. \ref{fig:closedform} shows the value of $\delta_1\delta_2$ versus the azimuth angle $\theta$ and elevation angle $\varphi$. One can observe that there exists truncation with a substantial region where the value of $\delta_1\delta_2$ approaches zero, indicating that the corresponding rotation angles fail to satisfy the constraints under the considered system setup. This highlights the importance of rotation optimization, as a considerable performance gain can be obtained by properly adjusting $\theta$ and $\varphi$. Furthermore, it is observed that all three schemes yield nearly identical optimal rotation angles $\bm{\Omega}$. Moreover, the elevation angle $\varphi$ is negative, which is expected since both the BS and the user are located below the IRS. The results demonstrate the effectiveness of the solution obtained by Algorithm \ref{Algo:single} and the closed form in achieving performance comparable to that by the exhaustive search method but with lower computational complexity.

\subsubsection{Performance Comparison for Rotatable IRS and Fixed Rotation IRS}
\begin{figure}[t]
	\centering
	\includegraphics[width=0.75\linewidth]{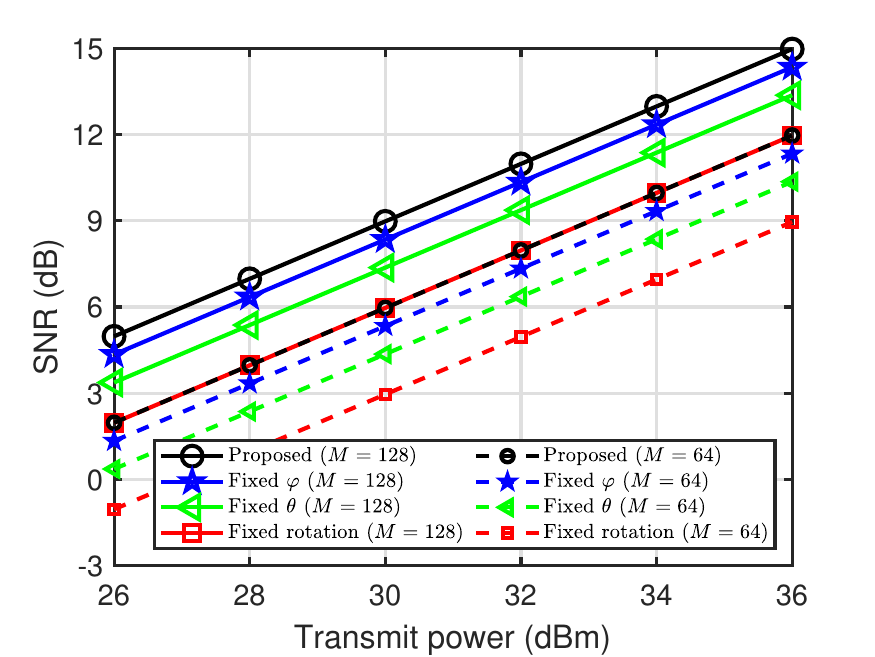}
	\vspace{-5pt}
	\caption{SNR versus transmit power.}
	\label{fig:M_single}
\end{figure}
In Fig. \ref{fig:M_single}, we plot the SNR versus the maximum transmit power at the BS for both $M = 128$ and $M = 64$. First, one can observe that the SNR of all the schemes monotonically increases with $P_\mathrm{t}$ and $M$ due to the higher beamforming gain provided by the BS. Second, it is observed that the proposed scheme outperforms other schemes, and the performance gap between the fixed rotation scheme and the proposed scheme is 3 dB. Moreover, we observe that the proposed scheme, even when deploying half the number of antennas at the BS, i.e., $M = 64$, is able to achieve almost the same performance as the fixed rotation scheme with $M = 128$. Based on \eqref{power}, which states that the received signal power is linearly proportional to the number of antennas, halving the antenna number introduces a 3 dB performance loss. The fact that the proposed scheme with $M=64$ can compensate for this theoretical loss and match the performance of the benchmark with $M=128$ demonstrates the effectiveness of the proposed scheme and highlights the importance of the IRS rotation.

\begin{figure}[t]
	\centering
	\includegraphics[width=0.75\linewidth]{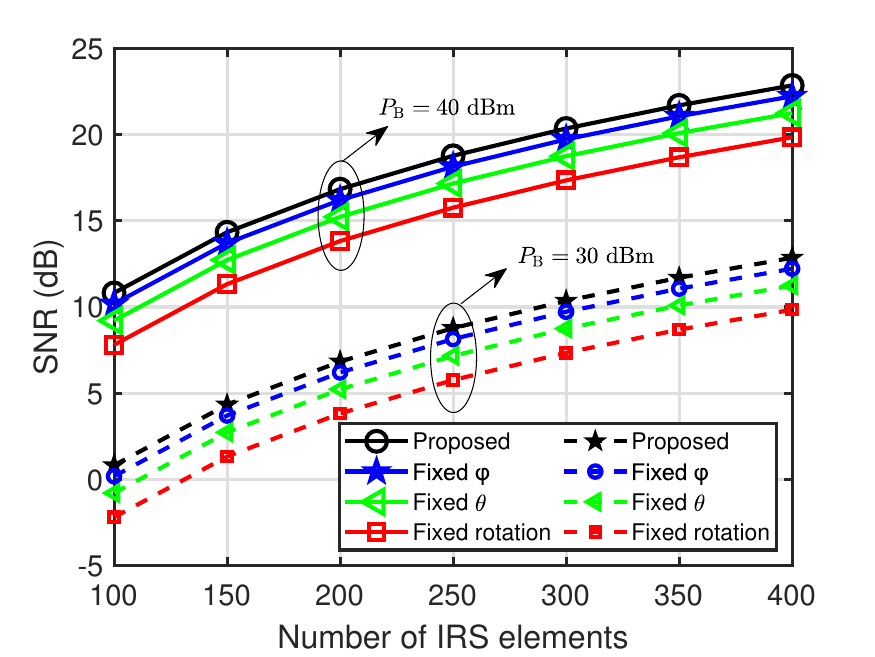}
	\vspace{-5pt}
	\caption{SNR versus the number of the IRS elements.}
	\label{fig:N_single}
\end{figure}
In Fig. \ref{fig:N_single}, we plot the SNR versus the number of the IRS elements for both $P_\mathrm{t}=30$ dBm and $P_\mathrm{t}=40$ dBm. It is observed that the SNR of all the schemes monotonically increases with $N$, which can be attributed to the fact that a larger number of reflecting elements contributes to higher passive beamforming gains, thereby enhancing the power of the received signal. Note that to achieve an SNR of 20 dB, the required number of IRS elements is reduced from 400 to 300 by replacing the fixed rotation scheme with the proposed scheme, which indicates that the proposed scheme substantially reduces the hardware requirements for the IRS configuration. The reason is that the proposed scheme provides larger effective reception and reflection factors that can effectively compensate for the performance degradation caused by the reduction in the number of the IRS elements.

\subsubsection{Performance Comparison for Rotatable IRS and M-IRS}
\begin{figure}[t]
	\centering
	\includegraphics[width=0.75\linewidth]{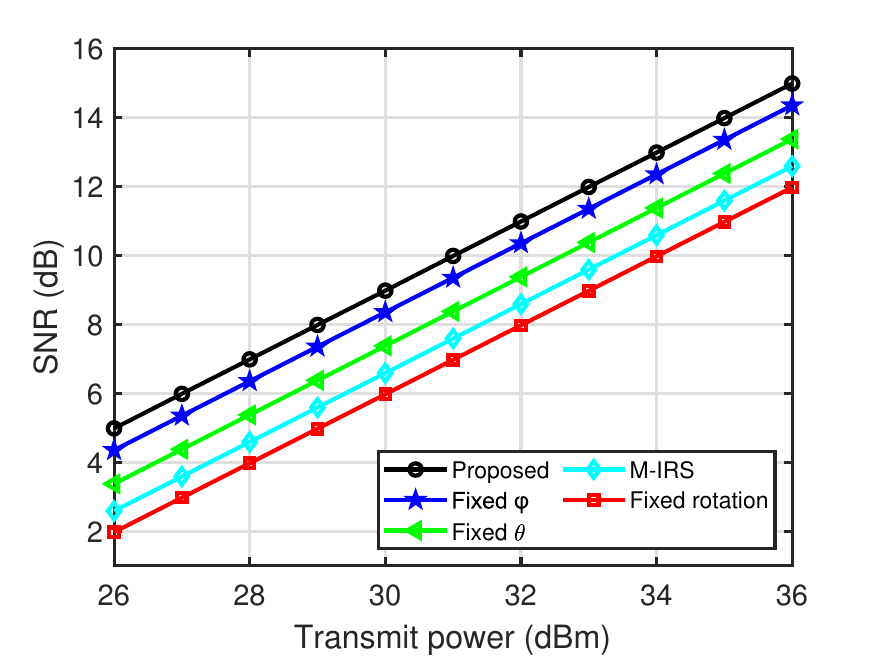}
	\vspace{-5pt}
	\caption{SNR versus transmit power.}
	\label{fig:RorM}
\end{figure}
Fig. \ref{fig:RorM} shows the SNR versus the transmit power. Compared to the fixed rotation scheme, where the IRS is unrotatable and unmovable, the M-IRS yields a higher SNR by additionally adjusting its placement. It is observed that the three schemes with rotatable IRS, even when only optimizing the azimuth angle or the elevation angle, perform better than that with M-IRS under different transmit power, thanks to the flexible rotation angles adjustment. Moreover, the SNR of the proposed scheme significantly outperforms other benchmark schemes. The result is expected since the proposed scheme can fully exploit the spatial design DoF of rotation, which highlights the potential of dynamically adjusting IRS rotation for single target location SNR enhancement.

\subsection{Area Coverage Enhancement}
\subsubsection{Performance Evaluation Under Locations within the Target Area}
\begin{figure}[t]
	\centering
	\includegraphics[width=0.75\linewidth]{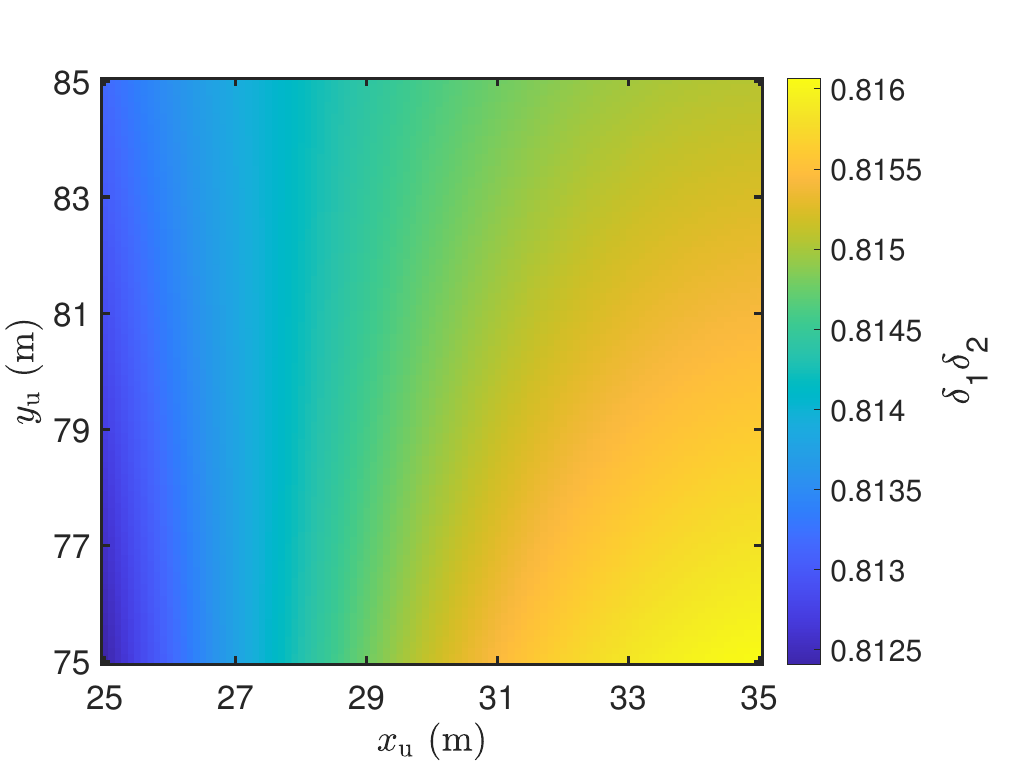}
	\vspace{-5pt}
	\caption{SNR at different locations within $\mathcal{A}$.}
	\label{fig:differentlocation}
\end{figure}
Fig. \ref{fig:differentlocation} shows the SNR at different locations within $\mathcal{A}$ for IRS-aided area coverage enhancement with the proposed scheme. It is observed that at any location within the target coverage area, an approximately equal value of $\delta_1\delta_2$ can be achieved, with all values being larger than 0.8124. Since $\delta_1\delta_2$ is a function of the reception and reflection factors of the IRS, the spatial uniformity across the target area indicates that the IRS not only efficiently captures the incident energy from the BS but also efficiently reflects it toward the user, irrespective of its specific locations. By optimizing the IRS rotation angles $\bm{\Omega}$, the system enables intelligent signal energy focusing and spatial distribution, which fully exploits the potential of the IRS in terms of SNR enhancement. The results strongly validate the effectiveness of the proposed scheme in enhancing the performance of the target coverage area.

\begin{figure}[t]
	\centering
	\includegraphics[width=0.75\linewidth]{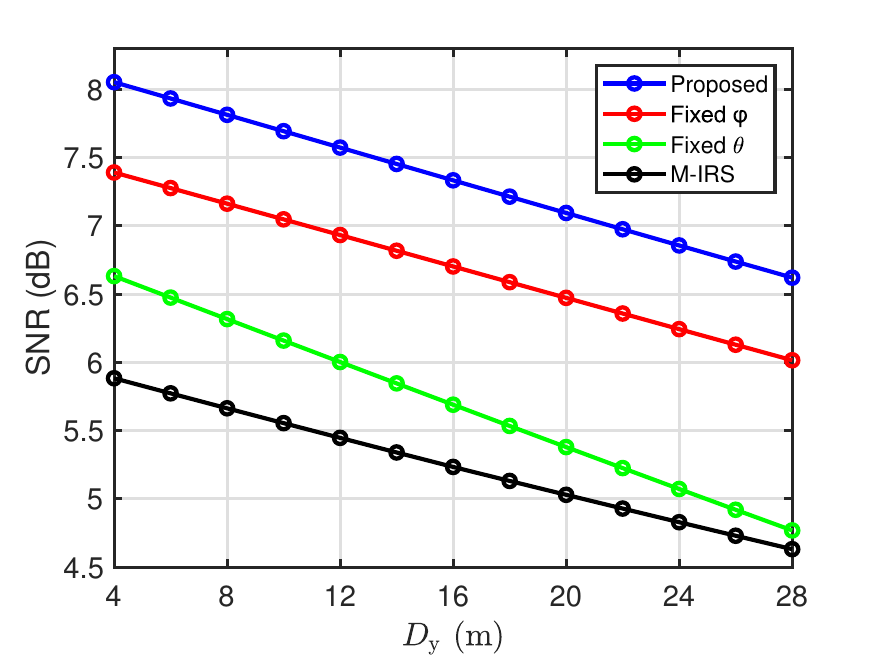}
	\vspace{-5pt}
	\caption{SNR versus the length of the target area, $D_y$.}
	\label{fig:SNR_Dy}
\end{figure}
In Fig. \ref{fig:SNR_Dy}, we plot the worst-case SNR versus the length of the target area, $D_y$. It is observed that all the schemes monotonically decrease as $D_y$ increases, with the fixed $\theta$ scheme exhibiting a more pronounced decline. For example, when $D_y$ increases from 4 m to 28 m, the SNR achieved by the fixed $\theta$ scheme drops from 6.6 dB to 4.7 dB, while the other three schemes experience a decrease of less than 1.5 dB. This is because the horizontal distance from the IRS to the edge of the target area becomes more significant, making the optimization of $\theta$ increasingly critical for target coverage enhancement. Nevertheless, the fixed $\theta$ scheme outperforms the M-IRS scheme across the entire range of $D_y$ in terms of SNR, particularly when the area length is relatively small. As the area size increases, it becomes increasingly challenging to achieve effective area coverage by optimizing only azimuth angle or elevation angle, due to the effective reflection constraints. Compared to the benchmark schemes, the proposed scheme achieves the best performance, which yields a significant performance gain over the other schemes. The results indicate that the IRS with the optimized rotation angles can fully exploit the spatial DoFs offered by the rotation design.

\subsubsection{Impact of the IRS's Altitude}
\begin{figure}[t]
	\centering
	\subfloat[SNR versus the IRS's altitude.]{
		\includegraphics[width=0.75\linewidth]{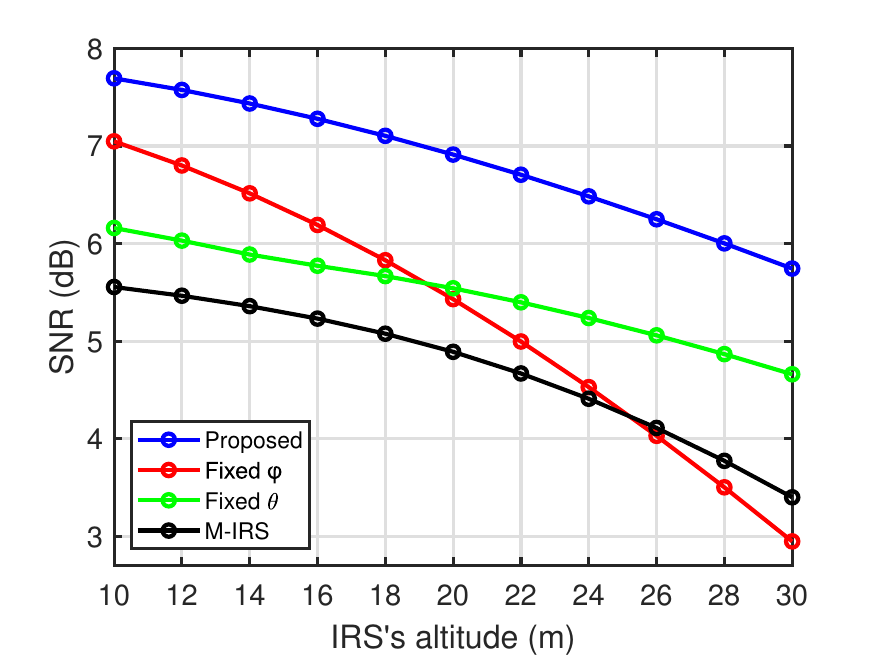}} \\ \vspace{-5pt}
	\subfloat[Optimized $\theta$ and $\varphi$ versus the IRS's altitude.]{
		\includegraphics[width=0.75\linewidth]{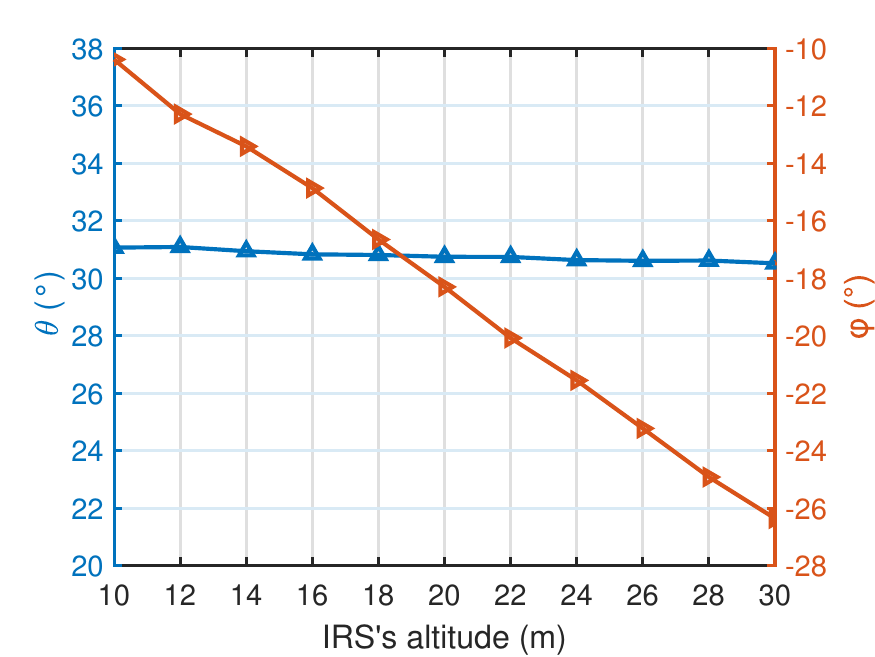}}
	\caption{Comparison on SNR and the optimized IRS rotation versus the IRS's altitude.}
	\label{fig:H}
\end{figure}
In Fig. \ref{fig:H}(a), we plot the worst-case SNR versus the IRS's altitude $z_\mathrm{c}$. One can observe that the schemes with optimized $\varphi$, i.e., the proposed scheme and the fixed $\theta$ scheme, outperform the M-IRS scheme under different $z_\mathrm{c}$. Moreover, we observe that the SNR of all the schemes monotonically decreases as $z_\mathrm{c}$ increases due to the increased path loss, with the fixed $\varphi$ scheme showing the sharpest decline. For example, as $z_\mathrm{c}$ increases from 18 m to 30 m, the SNR achieved by the fixed $\varphi$ scheme decreases from 5.8 dB to 2.9 dB. The gap between the schemes with and without optimized $\varphi$ becomes more pronounced as $z_\mathrm{c}$ increases. This is because the vertical distance from the IRS to both the BS and the user becomes more significant, and optimizing the elevation angle $\varphi$ becomes increasingly critical for SNR maximization. Nevertheless, the fixed $\varphi$ scheme performs better than the M-IRS scheme in a wide range of the IRS's altitude. In Fig. \ref{fig:H}(b), we plot the azimuth and elevation angle of the IRS optimized by Algorithm \ref{Algo:area} versus the IRS's altitude. It is observed that the absolute value of the elevation angle increases significantly with $z_\mathrm{c}$. In contrast, the optimized azimuth angle remains almost unchanged. It is because as the IRS's altitude increases, more vertical steering is required to aim at the $x$–$y$ plane, where the BS and user are located. The results demonstrate that optimizing $\varphi$ has a greater impact on improving the SNR than optimizing $\theta$, which are the same as Fig. \ref{fig:H}(a).

\section{Conclusion}
\label{Conclusion}
This paper has studied the rotatable IRS-aided wireless communication system under a practical angle-dependent channel model. Based on the developed model, the closed‑form SNR expression was derived to unveil the effects of IRS geometry, rotation, and electromagnetic factors modeling. PSO-based algorithms were developed to efficiently solve the optimization problem for SNR maximization. Numerical results demonstrated that the rotatable IRS outperformed the M‑IRS baseline in extending the coverage range with fewer BS antennas, transmit power, and reflecting elements. For the single target location case, it was shown that the proposed scheme achieves an approximately 3 dB SNR gain over the conventional fixed IRS scheme. For the area coverage case, adjusting the elevation angle of the IRS was shown to have a greater impact than tuning its azimuth angle on system performance at higher IRS altitudes. These findings provide useful guidelines for deploying the rotatable IRS in future wireless systems.

\section*{Appendix A: Proof of Proposition \ref{pro:1}}
Based on \eqref{p_n} and \eqref{p_b,m}, it follows that
\begin{align}
	t_{n,m} \!=\!& \left\| {{\bm{p}}_{{\mathrm{B,1}}}} \!+\! M_m^{\mathrm{c}}\tilde l{\bm{m}}_{\mathrm{B}}^{\mathrm{c}} \!+\! M_m^{\mathrm{r}}\tilde l{\bm{m}}_{\mathrm{B}}^{\mathrm{r}} \!-\! {{\bm{p}}_1} \!-\! N_n^{\mathrm{c}}l{{\bm{m}}_{\mathrm{c}}} \!-\! N_n^{\mathrm{r}}l{{\bm{m}}_{\mathrm{r}}} \right\| \nonumber\\
	=\!& \left\| \bm{d}_{1,1} \!+\! \Delta_{m,n} \right\|,n \in \mathcal{N},m \in \mathcal{M},
\end{align}
where $\bm{d}_{1,1} = {{\bm{p}}_{{\mathrm{B,1}}}} - {{\bm{p}}_1}$ and $\bm{\Delta}_{m,n} = M_m^{\mathrm{c}}\tilde l{\bm{m}}_{\mathrm{B}}^{\mathrm{c}} \!+\! M_m^{\mathrm{r}}\tilde l{\bm{m}}_{\mathrm{B}}^{\mathrm{r}} \!-\! N_n^{\mathrm{c}}l{{\bm{m}}_{\mathrm{c}}} \!-\! N_n^{\mathrm{r}}l{{\bm{m}}_{\mathrm{r}}}$ stands for the offset vector introduced by the array element indices. To facilitate the calculation, $\bm{\Delta}_{m,n}$ is decomposed w.r.t. $\bm{d}_{1,1}$. Assuming that the distance between the 1-th BS antenna and 1-th element of the IRS is larger than the size of the IRS, the component of $\bm{\Delta}_{m,n}$ parallel to $\bm{d}_{1,1}$ with magnitude $\bar \Delta$, contributes directly to the propagation distance, whereas its orthogonal component with magnitude $\tilde \Delta$, produces spatial deviation without affecting the distance magnitude. Then, it follows that
\begin{align}
	\bar \Delta &=\! \frac{\bm{\Delta}_{m,n} \! \cdot \! \bm{d}_{1,1}}{\|\bm{\Delta}_{m,n}\| \! \cdot \! \|\bm{d}_{1,1}\|} 
	\!=\! \frac{1}{t_{1,1}} \! \left( M_m^{\mathrm{c}}\tilde l{\bm{m}}_{\mathrm{B}}^{\mathrm{c}} \cdot \bm{d}_{1,1} \!+\! M_m^{\mathrm{r}}\tilde l{\bm{m}}_{\mathrm{B}}^{\mathrm{r}} \! \cdot \! \bm{d}_{1,1} \right.\nonumber\\
	&\left. - N_n^{\mathrm{c}}l{{\bm{m}}_{\mathrm{c}}} \cdot\bm{d}_{1,1} \!-\! N_n^{\mathrm{r}}l{{\bm{m}}_{\mathrm{r}}} \cdot\bm{d}_{1,1}\right),n \in \mathcal{N},m \in \mathcal{M}.
\end{align}
With $\varpi _{\mathrm{B}}^{\mathrm{c}}$, $\varpi _{\mathrm{B}}^{\mathrm{r}}$, ${\varpi _{\mathrm{c}}}$, and ${\varpi _{\mathrm{r}}}$, we have
\begin{align}
	\bar \Delta =& M_m^{\mathrm{c}}\tilde l\cos \varpi _{\mathrm{B}}^{\mathrm{c}} + M_m^{\mathrm{r}}\tilde l\cos \varpi _{\mathrm{B}}^{\mathrm{r}} \nonumber\\
	&- N_n^{\mathrm{c}}l\cos {\varpi _{\mathrm{c}}} - N_n^{\mathrm{r}}l\cos {\varpi _{\mathrm{r}}},n \in \mathcal{N},m \in \mathcal{M}.
\end{align}
According to the Pythagorean theorem, we have 
\begin{align}
	t_{n,m} = \sqrt{\left(t_{1,1} + \bar \Delta \right)^2+\tilde \Delta^2},n \in \mathcal{N},m \in \mathcal{M}.
\end{align}
Note that $t_{n,m}$ is a function with respect to $\tilde \Delta$ and thus we
use $t_{n,m}(\tilde \Delta)$ to represent $t_{n,m}$. When expanded by a Taylor series at $\tilde \Delta = 0$, $t_{n,m}(\tilde \Delta)$ can be represented as 
\begin{align}
	t_{n,m}(\tilde \Delta) \approx t_{n,m}(0) + t_{n,m}^{'}(0) \tilde \Delta.
\end{align}
Then, we can omit the second term since $\tilde \Delta$ is small, and approximate $t_{n,m}$ as
\begin{align}
	t_{n,m} & \approx t_{1,1} + \bar \Delta \nonumber\\
	&= {t_{1,1}} + M_m^{\mathrm{c}}\tilde l\cos \varpi _{\mathrm{B}}^{\mathrm{c}} + M_m^{\mathrm{r}}\tilde l\cos \varpi _{\mathrm{B}}^{\mathrm{r}} \nonumber\\
	&- N_n^{\mathrm{c}}l\cos {\varpi _{\mathrm{c}}} - N_n^{\mathrm{r}}l\cos {\varpi _{\mathrm{r}}} ,n \in \mathcal{N},m \in \mathcal{M}.
\end{align}
By substituting $t_{n,m}$ into \eqref{G}, the proof is completed.

\section*{Appendix B: Proof of Proposition \ref{pro:2}}
\label{Appendix A}
\begin{figure}[t]
	\centering
	\includegraphics[width=0.25\textwidth]{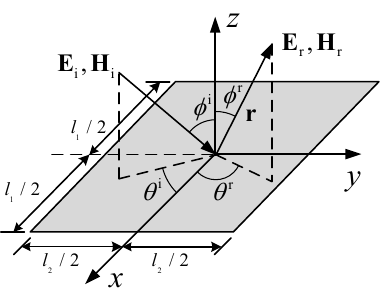}
	\vspace{-10pt}
	\caption{An example of electromagnetic wave scattering by a metal plate.}
	\label{fig:plane}
\end{figure}

We consider a perfectly conducting rectangular plate with dimensions $l_1 \times l_2$ and negligible thickness, located in the $x-y$ plane. A distant point source radiates linearly polarized waves with wave number $k=\frac{2 \pi}{\lambda}$. As shown in Fig. \ref{fig:plane}, let $\theta^\mathrm{i}$, $\phi^\mathrm{i}$, $\theta^\mathrm{r}$, and $\phi^\mathrm{r}$ denote the azimuth and elevation angles of the incident wave of the center of plane, as well as the elevation and azimuth angles of the reflected wave of the center of plane. The electric and magnetic field components of the incident wave is given by \cite{EM1,EM2}
\begin{align}
	{{\bm{E}}_{\mathrm{i}}} &= {E_i}\left( {{{\bm{e}}_{\mathrm{x}}}\sin {\theta ^{\mathrm{i}}} + {{\bm{e}}_{\mathrm{y}}}\cos {\theta ^{\mathrm{i}}}} \right) \nonumber\\
	&\times {e^{ - jk\left( { - x\sin {\phi ^{\mathrm{i}}}\cos {\theta ^{\mathrm{i}}} + y\sin {\phi ^{\mathrm{i}}}\sin {\theta ^{\mathrm{i}}} - z\cos {\phi ^{\mathrm{i}}}} \right)}},\\
	{{\bm{H}}_{\mathrm{i}}} &= \frac{{{E_i}}}{\mu }\left( {{{\bm{e}}_{\mathrm{x}}}\cos {\theta ^{\mathrm{i}}}\cos {\phi ^{\mathrm{i}}} - {{\bm{e}}_{\mathrm{y}}}\sin {\theta ^{\mathrm{i}}}\cos {\phi ^{\mathrm{i}}} - {{\bm{e}}_{\mathrm{z}}}\sin {\phi ^{\mathrm{i}}}} \right)\nonumber\\
	&\times {e^{ - jk\left( { - x\sin {\phi ^{\mathrm{i}}}\cos {\theta ^{\mathrm{i}}} + y\sin {\phi ^{\mathrm{i}}}\sin {\theta ^{\mathrm{i}}} - z\cos {\phi ^{\mathrm{i}}}} \right)}}, \label{H}
\end{align}
where $\mu$ is the characteristic impedance of the medium. Given the negligible thickness of the plane $Q$, the induced surface current density can be approximated
\begin{align}
	{{\bm{J}}_{Q}} &\approx 2{{\bm{e}}_{\mathrm{z}}} \times {{\bm{H}}_{\mathrm{i}}}{|_{z = 0, x=x',y=y'}} \nonumber\\
	&= \frac{{2{E_i}}}{\mu }\left( {{{\bm{e}}_{\mathrm{x}}}\cos {\theta ^{\mathrm{i}}}\cos {\phi ^{\mathrm{i}}} - {{\bm{e}}_{\mathrm{y}}}\sin {\theta ^{\mathrm{i}}}\cos {\phi ^{\mathrm{i}}}} \right) \nonumber\\
	&\times {e^{ - jk\left( { - x'\sin {\phi ^{\mathrm{i}}}\cos {\theta ^{\mathrm{i}}} + y'\sin {\phi ^{\mathrm{i}}}\sin {\theta ^{\mathrm{i}}}} \right)}}. \label{J}
\end{align}
Then, we have
\begin{align}
	&{J_{\mathrm{x}}} \!=\! \frac{{2{E_i}}}{\mu }\cos {\theta ^{\mathrm{i}}}\cos {\phi ^{\mathrm{i}}}{e^{ - jk ( - x'\sin {\phi ^{\mathrm{i}}}\cos {\theta ^{\mathrm{i}}} + y'\sin {\phi ^{\mathrm{i}}}\sin {\theta ^{\mathrm{i}}} )}},\\
	&{J_{\mathrm{y}}} \!=\! - \frac{{2{E_i}}}{\mu } \! \sin \theta ^{\mathrm{i}}\cos {\phi ^{\mathrm{i}}}{e^{ - jk ( - x' \! \sin {\phi ^{\mathrm{i}}}\cos {\theta ^{\mathrm{i}}} + y' \! \sin {\phi ^{\mathrm{i}}}\sin {\theta ^{\mathrm{i}}} )}}, \\
	&{J_{\mathrm{z}}} \!=\! 0.
\end{align}
Let $\bm{r}'=\left(x',y',0\right)^{T}$ denotes the position vector on the plate and $\zeta$ represents the angle between $\bm{r}$ and $\bm{r}'$. Then, we have $r'\cos \zeta = x'\sin {\theta ^{\mathrm{r}}}\cos {\phi ^{\mathrm{r}}} + y'\sin {\theta ^{\mathrm{r}}}\sin {\phi ^{\mathrm{r}}}$. By introducing an intermediate vector ${\bm{T}} = \int \int_Q {{{\bm{J}}_{\mathrm{Q}}}{e^{jkr'\cos \zeta}}dq'} $, it follows that
\begin{align}
	&{T_{\mathrm{x}}} = \int \int {{J_{\mathrm{x}}}{e^{jkr'\cos \zeta }}dx'dy'} ,\\
	&{T_{\mathrm{y}}} = \int \int {{J_{\mathrm{y}}}{e^{jkr'\cos \zeta }}dx'dy'} ,\\\
	&{T_{\mathrm{z}}} = 0,
\end{align}
when $z' = 0$. With the rectangular-to-spherical component transformation given by
\begin{align}
	\left[\! {\begin{array}{*{20}{c}}
			{{T_r}}\\
			{{T_\theta }}\\
			{{T_\phi }}
	\end{array}}\! \right] \!\!=\!\! \left[\! {\begin{array}{*{20}{c}}
			{\cos {\theta}\sin {\phi }}&{\sin {\theta }\sin {\phi }}&{\cos {\phi }}\\
			{\cos {\theta }\cos {\phi }}&{\sin {\theta }\cos {\phi }}&{ - \sin {\phi }}\\
			{ - \sin {\theta }}&{\cos {\theta }}&0
	\end{array}} \! \right] \!\! \left[\! {\begin{array}{*{20}{c}}
			{{T_{\mathrm{x}}}}\\
			{{T_{\mathrm{y}}}}\\
			{{T_{\mathrm{z}}}}
	\end{array}} \! \right]\!,
\end{align}
it follows that
\begin{align}
	&{T_\theta } \!=\! \frac{{2{E_i}{l_1}{l_2}}}{\mu } \! \cos {\phi ^{\mathrm{i}}}\cos {\phi ^{\mathrm{r}}}\cos ( {{\theta ^{\mathrm{i}}} \!+\! {\theta ^{\mathrm{r}}}} )\operatorname{sinc} ( X )\operatorname{sinc} ( Y ),\\
	&{T_\phi } \!=\! - \frac{{2{E_i}{l_1}{l_2}}}{\mu }\cos {\phi ^{\mathrm{i}}}\sin \left( {{\theta ^{\mathrm{i}}} \!+\! {\theta ^{\mathrm{r}}}} \right)\operatorname{sinc} ( X )\operatorname{sinc} ( Y ),
\end{align}
where $X = \frac{{\pi {l_1}}}{\lambda }\left( {\cos {\theta ^{\mathrm{i}}}\sin {\phi ^{\mathrm{i}}} + \sin {\theta ^{\mathrm{r}}}\cos {\phi ^{\mathrm{r}}}} \right)$ and $Y = \frac{{\pi {l_2}}}{\lambda }\left( {\sin {\theta ^{\mathrm{r}}}\sin {\phi ^{\mathrm{r}}} - \sin {\theta ^{\mathrm{i}}}\sin {\phi ^i}} \right)$. Assuming that the plate is a perfect electric conductor and in the far-field case, the scattered electric field magnitude is approximated by $\left| {{E_r}} \right| = \sqrt {{{\left| {{E_\theta }} \right|}^2} + {{\left| {{E_\phi }} \right|}^2}} $, where ${E_\theta } \approx - \frac{{jk{e^{ - jkr}}}}{{4\pi r}}\mu {T_\theta }$ and ${E_\phi } \approx - \frac{{jk{e^{ - jkr}}}}{{4\pi r}}\mu {T_\phi }$ with $r = \left|\bm{r}\right|$. As such, the ratio of the scattered electric field magnitude to the incident electric field magnitude is given by
\begin{align}
	\frac{{{\left|E_r\right|}}}{{{\left|E_{\mathrm{i}}\right|}}} = \frac{{{l_1}{l_2}}}{{\lambda r}}\cos {\phi ^{\mathrm{i}}}Z\operatorname{sinc} \left( X \right)\operatorname{sinc} \left( Y \right),
\end{align}
where $Z = \sqrt {{{{\cos }^2}{\phi ^{\mathrm{r}}}{{\cos }^2}\left( {{\theta ^{\mathrm{i}}} + {\theta ^{\mathrm{r}}}} \right) + {{\sin }^2}\left( {{\theta ^{\mathrm{i}}} + {\theta ^{\mathrm{r}}}} \right)}}$. After removing the path loss factor, i.e., $1/r$, and applying the spherical normalization used in radar cross section (RCS), i.e., $4\pi$, the reflection coefficient of the metal plate without the IRS phase shift is given by
\begin{align}
	\bar \eta = \sqrt {\frac{{4\pi l_1^2l_2^2}}{{{\lambda ^2}}}} \cos {\phi ^{\mathrm{i}}}Z\operatorname{sinc} \left( X \right)\operatorname{sinc} \left( Y \right).
\end{align}
Thus, the proof is completed.

\section*{Appendix C: Proof of Proposition \ref{small}}
Given that both squared terms are non-negative, assuming that the objective function of problem \eqref{pro:power_small} can reach its maximum value of 1, i.e., $\delta _1 = 1$, then we have $\cos^2 {{ \phi }^{\mathrm{i}}} = 1$ and ${{{\cos }^2}{{ \phi }^{\mathrm{r}}}{{\cos }^2}\left( {{{ \theta }^{\mathrm{i}}} + {{ \theta }^{\mathrm{r}}}} \right) + {{\sin }^2}\left( {{{ \theta }^{\mathrm{i}}} + {{ \theta }^{\mathrm{r}}}} \right)} = 1$.
Two feasible solutions that satisfy the above conditions are given by
\begin{align}
	\begin{cases}
		{{{ \phi }^{\mathrm{i}}} = 0},\\
		{{{\theta }^{\mathrm{i}}} = 0},\\
		{{{\theta }^{\mathrm{r}}} = \pi/2},
	\end{cases}
	\text{and }
	\begin{cases}
		{{{ \phi }^{\mathrm{i}}} = 0},\\
		{{{\theta }^{\mathrm{i}}} = 0},\\
		{{{\theta }^{\mathrm{r}}} = 3\pi/2}.
	\end{cases}
\end{align}
With both solutions, we have $\theta^{\mathrm{i}} = 0$. Substituting $\phi^{\mathrm{i}} = 0$ into \eqref{theta_i} yields the following equation for the optimized azimuth angle: ${x_{\mathrm{B}}}\sin \theta^{\mathrm{opt}} + \left( {{y_{\mathrm{B}}} - {y_{\mathrm{c}}}} \right)\cos \theta^{\mathrm{opt}} = 0$.
Thus, we have $\theta^{\mathrm{opt}} = \arctan \left( { - \frac{{{y_{\mathrm{B}}} - {y_{\mathrm{c}}}}}{{{x_{\mathrm{B}}}}}} \right)$.
Given $\phi^{\mathrm{i}} = 0$, by substituting $\theta^{\mathrm{opt}}$ into \eqref{phi_i}, the equation for the optimized elevation angle is given by $\sqrt {x_{\mathrm{B}}^2 + {{( {{y_{\mathrm{B}}} - {y_{\mathrm{c}}}} )^2}}} \cos \varphi^{\mathrm{opt}} - {z_{\mathrm{c}}}\sin \varphi^{\mathrm{opt}} = \sqrt {x_{\mathrm{B}}^2 + {{( {{y_{\mathrm{B}}} - {y_{\mathrm{c}}}})^2}} + z_{\mathrm{c}}^2}$.
Thus, we have $\varphi^{\mathrm{opt}} = \arctan \left( \frac{{ - {z_{\mathrm{c}}}}}{{\sqrt {x_{\mathrm{B}}^2 + {{( {y_{\mathrm{B}}} - {y_{\mathrm{c}}})^2}}} }} \right)$.
By substituting $\theta^{\mathrm{opt}}$ and $\varphi^{\mathrm{opt}}$ into \eqref{theta_r}, we have
\begin{align}
	\theta^\mathrm{r} \!\!=\! \arctan \! \frac{{( {{x_{\mathrm{B}}}( {{y_{\mathrm{U}}} \!-\! {y_{\mathrm{c}}}} ) \!-\! {x_{\mathrm{U}}}( {{y_{\mathrm{B}}} \!-\! {y_{\mathrm{c}}}} )} )\sqrt {x_{\mathrm{B}}^2 \!\!+\!\! {( {{y_{\mathrm{B}}} \!-\! {y_{\mathrm{c}}}} )^2} \!+\! z_{\mathrm{c}}^2} }}{{{z_{\mathrm{c}}}( {x_{\mathrm{B}}^2 \!+\! {( {{y_{\mathrm{B}}} \!-\! {y_{\mathrm{c}}}})^2} \!-\! {x_{\mathrm{B}}}{x_{\mathrm{U}}} \!-\! ( {{y_{\mathrm{U}}} \!-\! {y_{\mathrm{c}}}} )( {{y_{\mathrm{B}}} \!-\! {y_{\mathrm{c}}}} )} )}}.
\end{align} 
Assuming that $\frac{{\left( {{x_{\mathrm{B}}}\left( {{y_{\mathrm{U}}} - {y_{\mathrm{c}}}} \right) - {x_{\mathrm{U}}}\left( {{y_{\mathrm{B}}} - {y_{\mathrm{c}}}} \right)} \right)\sqrt {x_{\mathrm{B}}^2 + {{\left( {{y_{\mathrm{B}}} - {y_{\mathrm{c}}}} \right)}^2} + z_{\mathrm{c}}^2} }}{{{z_{\mathrm{c}}}\left( {x_{\mathrm{B}}^2 + {{\left( {{y_{\mathrm{B}}} - {y_{\mathrm{c}}}} \right)}^2} - {x_{\mathrm{B}}}{x_{\mathrm{U}}} - \left( {{y_{\mathrm{U}}} - {y_{\mathrm{c}}}} \right)\left( {{y_{\mathrm{B}}} - {y_{\mathrm{c}}}} \right)} \right)}} \gg 1$, $\theta^{\mathrm{r}}$ can be approximated as $\pi/2$, which completes the proof.

\bibliographystyle{IEEEtran}
\bibliography{refs.bib} 

\end{document}